\documentclass{article}
 \pdfoutput=1
 
\usepackage{geometry}
\usepackage{graphicx}

\usepackage[table]{xcolor}
\usepackage{subfigure}
\usepackage{float}
\usepackage{booktabs}
\usepackage{lpic}
\usepackage{verbatim}
\usepackage{amsmath}
\usepackage{color}
\usepackage{rotating}
\usepackage{nomencl}
\usepackage{makecell}

\usepackage{bm}
\usepackage{authblk}
\usepackage{multirow,array}
\date{}

\title{Parameter based design of a twin-cylinder wave energy converter for real sea-states}

\author{Dali Xu\footnote{Present address: College of Ocean Science and Engineering, Shanghai Maritime University, 1550 Hai Gang Da Dao, 201306 Shanghai, China}}
\author{Raphael Stuhlmeier} 
\author{Michael Stiassnie}
\affil{Faculty of Civil and Environmental Engineering,\\
Technion-Israel Institute of Technology, 32000 Haifa, Israel}

\begin{document}

\maketitle

\begin{abstract}
We discuss the hydrodynamics of a wave energy converter consisting of two vertically floating, coaxial cylinders connected by dampers and allowed to heave, sway and roll. This design, viable in deep water and able to extract energy independent of the incident wave direction, is examined for monochromatic waves as well as broad-banded seas described by a Pierson Moskowitz spectrum. Several possible device sizes are considered, and their performance is investigated for a design spectrum, as well as for more severe sea states, with a view towards survivability of the converters. In terms of device motions and captured power, a quantitative assessment of converter design as it relates to survival and operation is provided. Most results are given in dimensionless form to allow for a wide range of applications.
\end{abstract}

\section{Introduction}
\label{sec:Introduction}
Cylindrical floating bodies are ubiquitous in ocean engineering. In part, this is due to the fact that the most accessible cases in the linear theory of wave-structure interaction conform to the geometries for which the field equation -- Laplace's equation -- is separable. In three spatial dimensions, the most straightforward such geometry is the circular cylinder.
Due to the accessibility, and wealth of prior studies (a brief review of which is given later in the Introduction), the heaving truncated cylinder is an exceedingly popular model for a wave energy converter (WEC), able to absorb energy from waves independent of direction.  Due to this very ubiquity, we feel the design parameters -- in this case limited to cylinder size and damping -- deserve a separate, careful investigation. Furthermore, we aim to extend prior work by considering all three modes of motion available to an axisymmetric body.

The intention of this study is to present a rather comprehensive account of the hydrodynamics of a WEC based on relative motion of two cylindrical bodies, which are allowed to move in heave, as well as sway and roll. This entails an investigation of different WEC sizes and damping coefficients and their performance in wind-generated sea surface waves given by a Pierson-Moskowitz (PM) spectrum. The spectral description of the sea surface allows one to derive values for the \emph{significant displacements} of the WEC, and thus add a measure of survivability to the design considerations. Finally, an illustrative grading system can be devised to categorize the various performance metrics of the self-reacting WECs.

Following the work of Falc\~ao \cite{Falcao2010} we classify WECs into three different types based on different working principles: oscillating water columns, overtopping devices, and oscillating body devices. For the most part, offshore devices fall into the last category. Furthermore, since energy production by oscillating bodies is governed by the wave-induced motion, the dimensions of such WECs are constrained significantly by the incident wavelengths, and will likely be roughly similar under given conditions. Subsequently, oscillating body devices may be further subdivided into those reacting against a fixed frame of reference (either the sea bed or a bottom fixed structure), and those reacting against one or more floating bodies, called \emph{wave-activated bodies} or \emph{self-reacting devices.} These last devices may be installed in deep water, where the large distance between the sea-bed and the surface might otherwise be prohibitive. The mooring system for such devices has the sole role of counteracting drift and current forces, allowing the weight of moorings and anchors to be relatively small (see \cite{Cerveira2013} and references therein).

We consider vertically floating cylinders, for which a rich literature exists. The radiation problem in heave only was addressed by Ursell in 1949 \cite{Ursell1949}, and the scattering problem by Dean and Ursell \cite{Ursell1959}.  Miles and Gilbert \cite{Miles1968} later employed a variational approximation to provide the far field potential for scattering by a circular dock, along with the lateral forces on the dock. However, their results were subsequently found to contain several inaccuracies, in particular in their calculations of the radiation forces. This prompted Garrett\cite{Garrett1971} to take up the problem afresh, and establish the scattering forces for a circular dock. Subsequently, Black et al.\cite{Black1971} revisited the application of variational methods to the radiation and scattering problem by several cylindrical geometries, employing Haskind's theorem to give the wave forces.
This latter, variational approach did not yield the added mass and damping coefficients. Hence in 1981, Yeung\cite{Yeung1981} studied the radiation problem of a vertical cylinder floating on the water surface and undergoing the combined motions of heave, sway and roll, and obtained these hydrodynamic coefficients. More recently, Bhatta\cite{Bhatta2007} also gave the added mass and damping coefficient of a vertical cylinder undergoing heave motion, in terms of the two dimensionless ratios characterizing the problem (depth to radius and draft to radius).
While prior work had focused on the finite depth case, in 2013, Finnegan et al \cite{Finnegan2013} treated by means of an analytical approximation due to Leppington the forces on a truncated vertical cylinder in water of infinite depth.

In the context of wave energy, the consideration of floating cylinders as models of WECs goes back at least to Berggren \& Johansson \cite{Berggren1992}, who  approximated a device described by Hagerman by two floating, axisymmetric cylinders oscillating in heave, albeit without any considerations of captured power. More recently, Garnaud and Mei \cite{Garnaud2009} revisited the single buoy with the intention of studying it in densely packed arrays, giving the captured power for buoys hanging from a large frame. This floating point absorber was also employed, e.g.\ by Child and Venugopal \cite{Child2010} in their discussion of optimization of WEC arrays, or by Borgarino et al \cite{Borgarino2012} as a generic model to investigate wave interaction effects. Similarly, Teillant et al \cite{Teillant2012} employ an axisymmetric, heaving 2-body device for their study of WEC economics, without detailed hydrodynamic considerations. Meanwhile, a slightly different construction of the floating body was considered by Engstr{\"{o}}m et al \cite{Engstrom2009}, who added a sphere under the floating cylinder.  This two-body configuration of floating cylinder and submerged sphere was then assumed connected to the sea bed by a generator, and its performance analyzed. Zheng et al \cite{Zheng2005}, in a generalization of Berggren \& Johansson to three modes of motion, considered the hydrodynamics of two unconnected, coaxial floating cylinders, again without considering power capture. The power capture for a heave-only two-cylinder WEC was recently obtained for attacking monochromatic incident waves by Wu et al \cite{Wu2014}, albeit with a rather terse discussion of their results. 

We combine features of several previous authors, and consider the novel case of two floating cylinders with an idealized power take--off (PTO), represented by a linear damper of constant characteristics, whose optimization is part of the design procedure.\footnote{While studies on PTO control show a promising potential for enhancing performance, particularly for devices with a narrow-banded natural response, practical and robust applications must still be developed (see Hong et al \cite{Hong2014}).}. We undertake our parametric study with an eye towards applications, and thus consider irregular waves in the form of a PM spectrum (see e.g.\ recent work on optimizing a floating box-barge under irregular waves by \ B\'{o}dai \& Srinil \cite{Bodai2015}) While scatter diagrams may be available for some sites where an assessment of the wave resource has been carried out, where this is not the case estimates based on wind speed will need to be made. To this end, we present our data nondimensionalized on the basis of wind speed, which uniquely determines the PM spectrum. Values of significant wave height and peak period may be readily derived therefrom, and the data recast in these terms if desired.

The paper is organized as follows: in Section \ref{sec:physical preliminaries} we present the physical set-up of the problem. This consists in presenting the twin cylinder WEC and characterizing its geometry, and subsequently presenting the PM spectra for design and survivability considerations. In Section \ref{sec:governing equations} we present, very briefly, the basic mathematical formulation of the governing equations and sketch the solution procedure. Subsequently, we employ the hydrodynamic coefficients and forces found from solving the equations of Section \ref{sec:governing equations} to characterizing WEC design under monochromatic waves in Section \ref{sec:design of the wec}, and under irregular waves given by a Pierson-Moskowitz spectrum in Section \ref{section: irregular design}. A discussion of these results with a view to applications is given in Section \ref{section:discussion}, which is subdivided into Section \ref{subsec:power capture} on power capture, Section \ref{subsec: survivability} on survivability, and Section \ref{subsec:grading} which presents a synthesis of the preceding sections. Finally, Section \ref{sec:conclusions} presents some concluding remarks and perspectives.

\section{Physical preliminaries}
\label{sec:physical preliminaries}
\subsection{Geometry}

\begin{figure}[h!]
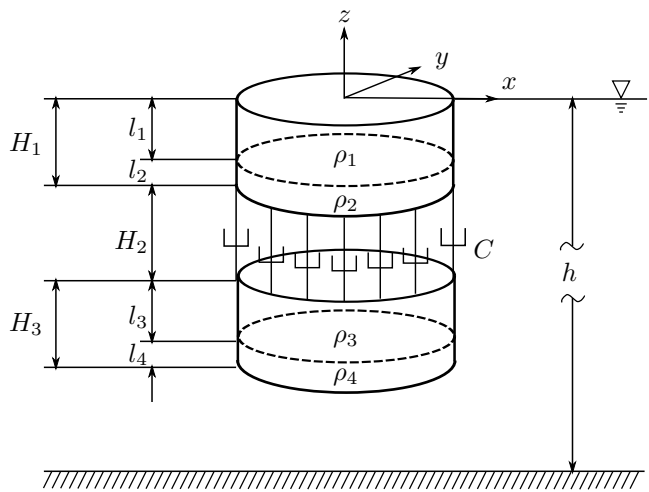

\begin{center}
\begin{lpic}[l(0mm),r(0mm),t(2mm),b(1mm)]{TWINCYLINDERS_NEW(0.2)}
\lbl[r]{0,110;$H_3$}
\lbl[r]{0,230;$H_1$}
\lbl[r]{70,165;$H_2$}
\lbl[r]{70,212;$l_2$}
\lbl[r]{70,240;$l_1$}
\lbl[r]{70,115;$l_3$}
\lbl[r]{70,90;$l_4$}
\lbl[r]{210,220;$\rho_1$}
\lbl[r]{210,190;$\rho_2$}
\lbl[r]{210,100;$\rho_3$}
\lbl[r]{210,75;$\rho_4$}
\lbl[r]{355,145;$h$}
\lbl[r]{205,315;$z$}
\lbl[r]{315,270;$x$}
\lbl[r]{270,285;$y$}
\lbl[r]{300,160;$C$}
\end{lpic}
\end{center}
\caption{Schematic depiction of the WEC geometry.}
\label{fig:geometry}
\end{figure}

The geometry and basic parameters of the twin-cylinder WEC are depicted in Fig \ref{fig:geometry}. The $Oxy$ plane is the still water surface and the $z$-axis points upwards. ($r, \theta$) are polar coordinates in the horizontal plane, such that $x = r\cos \theta$ and $y = r \sin \theta.$ The upper cylinder floats on the water surface with a draft $H_1$. To provide for flotational stability, it is important to note that the mass of this cylinder is not uniformly distributed, but is divided into two parts with drafts $l_1$ and $l_2$ and densities $\rho_1$ and $\rho_2$, respectively. The lower cylinder is entirely submerged with a draft $H_3$, and like the upper cylinder is assumed divided into two parts with densities $\rho_3$ and $\rho_4$ and drafts $l_3$ and $l_4$, respectively. The distance between the two cylinders is $H_2$. Both of them have the same radius $R$, and the water depth $h$ is taken to be very large compared to the attacking wave length, with the intention of approximating deep-water conditions.

As shown in Fig.\ref{fig:geometry}, the two cylinders are connected by a continuously distributed dashpot, which connects the upper edge of the lower cylinder with the lower edge of the upper cylinder at $r=R$. The integrated dashpot coefficient is $C$, which results in a dashpot coefficient per length  $\frac{C}{2\pi R}$. The dashpot is considered to represent a PTO, which generates energy from both the relative heave and roll motion of the cylinders.\footnote{Due to the small effect of sway motions, it is not necessary to consider power take-off in the sway mode for this device geometry.}

Since the two cylinders are axisymmetric, only these three modes are studied. The heave and sway motions will give rise to relative motions in $z$ and $x$ directions, respectively. For waves propagating in the $x-$direction, the two cylinders roll around the $y$-axis in the mean free surface ($z=0$), yielding a relative angle about this axis.

This formulation of the problem leaves us with thirteen parameters ($\{H_i \mid i\in \{1,2,3\}\}, \{l_j,\rho_j \mid j \in \{1,2,3,4\} \}, R,\text{ and } C$) characterizing the WEC. Before proceeding, we will make several restrictions to ensure that the problem remains manageable; nevertheless, we shall see that a wealth of interesting phenomena and properties of the WEC are still accounted for.

For simplicity, we will take the drafts and distance between the cylinders identical to their radius, and denote the single size parameter by $q$, i.e.
\begin{equation} \label{eq: cylinder sizes}
H_1=H_2=H_3=R \equiv q .
\end{equation}
For the density distribution of the cylinders, we shall assume
\begin{equation}
\rho_1=\rho_3=\frac{3}{4}\rho, \;\; \rho_2=\rho_4=\frac{3}{2}\rho,\;\; l_1=l_3=2l_2=2l_4=\frac{2}{3}q,\label{casestudy}
\end{equation}
where $\rho$ is the density of the water. 
Thus, the design problem  is reduced to two parameters, a size $q$ and dashpot coefficient $C$, whose interplay with incoming waves of certain frequencies is the issue at hand. We shall see that suspending the lower cylinder at a depth $2q$ below the still water surface has the desired effect of rendering its motion rather small, and thus creating a relatively stable point for the upper cylinder to react against.

There are several reasonable criteria which may govern the design of a WEC. Evidently, the WEC should capture as much of the incoming wave energy as possible. At the same time, as economic viability is the prime driver behind wave energy technology, the costs -- which may be assumed to scale with device size -- should be kept low; in practical terms, this means that device size should be kept small. Competing with this are concerns over the survivability of the converter, which dictate that displacements of the WEC not be too large under severe conditions, favoring larger devices. We shall return to these issues in detail in later sections.

\subsection{The Pierson Moskowitz spectrum}
\label{subsect:the pm spectrum}
One of the most common descriptions of a sea-state for engineering purposes is the unidirectional Pierson Moskowitz (PM) spectrum, here given as a function of wavenumber $k:$ 
\begin{equation}
S(k)=\frac{0.00405}{k^3}\exp{\left\{-0.55411\frac{g^2}{U^4 k^2}\right\}},\label{pm2}
\end{equation}
where $U$ is the mean wind speed at a height of $10$ m above the mean surface level, and $g$ is the gravitational acceleration. This empirically derived formula gives the energy distribution for wind waves in deep water, and differs from the JONSWAP spectrum only by the addition of a spectral--peak enhancement factor. 

This spectrum (\ref{pm2}) readily yields a number of important values associated with the sea-state:
\begin{align}
& H^{(1/3)}={0.24181 U^2}/{g},  \label{eq:PM to Hs}\\ \label{eq:PM to peak wavenum}
& k_p={0.66570 g}/{U^2}, 
\end{align}
where $ H^{(1/3)}$ is the significant wave height and $k_p$ is the wave number of the spectral peak for a given wind speed $U$. This makes it easy to present subsequent results in an alternative form when desired.
A monochromatic wave with wavenumber $k_p$ and the same wave energy density as the PM spectrum will have an amplitude
\begin{equation} \label{eq:monochrom ampl}
a_0(k_p) = {0.08549 U^2}/{g}.
\end{equation}

For subsequent illustration it will be necessary to have some concrete, physical examples, which means specifying a sea-state via a wind speed value $U.$ Our design conditions (denoted by a subscript $d$) will correspond to a wind speed $U_d = 10$ m/s, while we will consider two ``severe states" (denoted by subscripts $s1$ and $s2$) with regard to the survivability, corresponding to wind speeds $U_{s1} = 15$ m/s and $U_{s2} = 20$ m/s. These are summarized in Table \ref{SwSs}.

\renewcommand{\arraystretch}{1.2}
\begin{table}[!htb]
\caption{The wind speed $U,$ significant wave height $H^{(1/3)},$ peak wavenumber $k_p,$ and peak wavelength $\lambda_p = 2 \pi/k_p$ associated with PM spectra used for design and survivability considerations.} % title name of the table
\centering % centering table
\begin{tabular}{@{} l c c c @{}}
  \toprule
  &$S_d$&$S_{s1}$&$S_{s2}$\\
  \cline{2-4}
  
  $U \text{ (m/s)}$	&10&15&20\\
$H^{(1/3)}\text{ (m)}$&2.47& 5.55&9.87 \\
$k_p\text{ (1/m) }$	&0.065 & 0.029 &0.016	  \\
$\lambda_p\text{ (m)}$	&96.30 & 216.67 &385.19	  \\
\bottomrule
  \end{tabular}
\label{SwSs}
\end{table}

\section{Governing equations}
\label{sec:governing equations}

Our approach to solving the wave-structure problem for the twin-cylinder WEC relies on domain decomposition and eigenfunction expansion methods (in the context of floating cylinders, see Black \& Mei \cite{Black1970}, who give a comprehensive description of the method, or more generally, Linton \& McIver \cite{Linton2001}, or Zheng et al \cite{Zheng2005} for a recent application to floating cylinders).  As the full formulation is rather lengthy, we only indicate the most important equations, and refer the interested reader to work cited above.

The fluid is assumed to be incompressible and inviscid, and the flow irrotational. Introducing a velocity potential $\Phi(r,\theta,z,t),$ and assuming periodic motion of frequency $\omega,$ the potential is separated into the spatial and temporal parts,
\begin{equation}
\Phi(r,\theta,z,t)=\phi(r,\theta,z)e^{i\omega t},\label{sep}
\end{equation}
where $\phi(r,\theta,z)$ satisfies the Laplace equation:
\begin{equation}\label{ge:total}
\phi_{rr}+\frac{1}{r}\phi_{r}+\frac{1}{r^2}\phi_{\theta\theta}+\phi_{zz}=0,
\end{equation}
subject to the linearized boundary conditions on the free surface  $z = 0$ and on the bed $z = -h$:
\begin{align}\label{bc:total}
& \phi_z-\sigma\phi = 0, \text{ on } z=0,\;\;r>R, \\
& \phi_z = 0, \text{ on } z=-h, \label{bc:total:2}
\end{align}
where $\sigma=\omega^2/g.$

At the interface between structure and fluid, the normal velocity of the structure must equal that of the adjacent fluid particles, written in terms of the potential \eqref{sep}:
\begin{equation} \label{eq:body boundary condition - general}
\frac{\partial \Phi}{\partial n} = V_n,
\end{equation}
where $V_n$ is the component of the structure's velocity in the direction of the outward pointing normal vector $n,$ which may be applied at the equilibrium surface under the assumptions of linearity. Owing to this very linearity, we continue with a decomposition of the problem into two parts: one due to the waves ($\phi_S$) scattered from the structure (which is assumed fixed) by the incident wave field, and one due to the waves ($\phi_R$) radiated by the motion of the structure, such that $\phi = \phi_S + \phi_R.$ $\phi_S$ is decomposed further into the potential due to the incident wave $\phi_I$ and that due to the waves diffracted from the fixed structure $\phi_D,$ where 
\begin{equation}
\frac{\partial \phi_D}{\partial n} = -\frac{\partial \phi_I}{\partial n} \text{ on the body surface } S.
\end{equation}
The remaining radiated part of the potential $\phi_R$ must then satisfy \eqref{eq:body boundary condition - general}, where the normal velocities are to be determined from the equations of motion. We shall consider an incident monochromatic wave with amplitude $a_0,$ so that $\phi_I$ is known \emph{a priori.}

Introducing the as-yet unknown displacements of the upper ($j=1$) and lower ($j=2$) cylinder for the three modes of motion
\begin{eqnarray}
\zeta_{zj}&=&\zeta_{zj0}e^{i\omega t}\;\;\text{for heave},\\
\zeta_{xj}&=&\zeta_{xj0}e^{i\omega t}\;\;\text{for sway},\\
\theta_{j}&=&\theta_{j0}e^{i\omega t}\;\;\text{for roll},
\end{eqnarray}
where $\zeta_{zj0}$, $\zeta_{xj0}$ and $\theta_{j0}$ are the complex amplitudes of the corresponding displacements, we can write the boundary condition \eqref{eq:body boundary condition - general} for the spatial part of the total potential $\phi$ in the following form
\begin{align}
& \phi_z = i\omega\zeta_{z10} - i\omega \theta_{10} r\cos\theta, \quad &\text{ on } z=-H_1, \, r<R,\label{bc:total:3} \\
& \phi_z = i\omega\zeta_{z20} -i\omega \theta_{20}r\cos\theta, \quad &\text{ on } z=-(H_1+H_2), \, z=-(H_1+H_2+H_3), \, r<R \label{bc:total:4} \\
& \phi_r = i\omega \zeta_{x10}\cos\theta -i\omega \theta_{10}(z_0-z)\cos\theta, \quad &\text{ on } -H_1<z<0, \, r=R,\label{bc:total:5} \\
& \phi_r = i\omega \zeta_{x20}\cos\theta -i\omega \theta_{20}(z_0-z)\cos\theta, \quad &\text{ on } -(H_1+H_2+H_3)<z<-(H_1+H_2), \, r=R,\label{bc:total:6}
\end{align}
where \eqref{bc:total:3} is posed on the bottom of the upper cylinder, \eqref{bc:total:4} on the top and bottom of the lower cylinder, \eqref{bc:total:5} on the sides of the upper cylinder, and \eqref{bc:total:6} on the sides of the lower cylinder.
These conditions are supplemented by Sommerfeld's radiation condition, requiring the diffracted and radiated waves to be outgoing as $r \rightarrow \infty.$
\begin{figure}[h!]
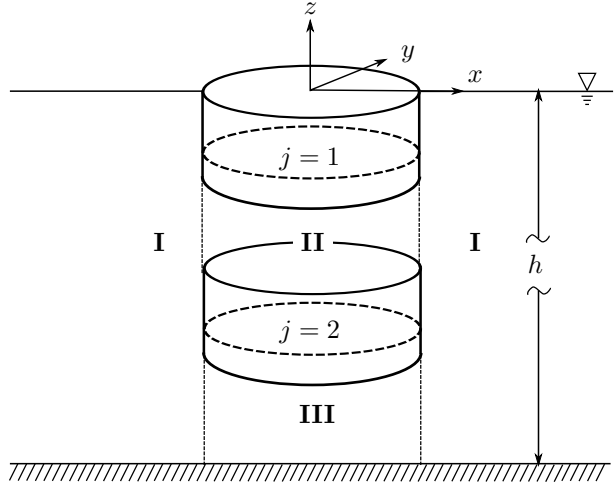

\begin{center}
\begin{lpic}[l(0mm),r(0mm),t(2mm),b(1mm)]{Subdomains_NEW(0.2)}
\lbl[c]{200,215;$j=1$}
\lbl[c]{200,100;$j=2$}
\lbl[r]{355,145;$h$}
\lbl[r]{205,315;$z$}
\lbl[r]{315,270;$x$}
\lbl[r]{270,285;$y$}
\lbl[Wc]{205,160;$\textbf{II}$}
\lbl[c]{205,45;$\textbf{III}$}
\lbl[c]{100,160;$\textbf{I}$}
\lbl[c]{310,160;$\textbf{I}$}
\end{lpic}
\end{center}
\caption{Domain decomposition for the twin-cylinder problem.}
\label{fig:subdomains}
\end{figure}
Due to the configuration of two axisymmetric floating cylinders, we must consider three fluid regions, one between the coaxial cylinders (region II), one between the lower cylinder and the bed (region III), and one outside the vertical extension of the cylinders (region I), as depicted in Figure \ref{fig:subdomains}. Subsequently the scattering problem is divided into three problems, one in each subdomain, and the radiation problem for each of the three modes of each of the two cylinders is divided into three problems. The reader may appreciate the effort involved in keeping track of, solving, and subsequently matching solutions, of 21 problems for the potentials involved. These potentials are then applied in calculating the forces on the two cylinders, in the form of pressures from the surrounding fluid.

The full expressions for the exciting, hydrodynamic, and hydrostatic forces are lengthy and will not be given. We note only that we have found excellent agreement between our results and published work  \cite{Garrett1971,Yeung1981,Berggren1992,Zheng2005,Bhatta2007,Garnaud2010,Bhatta2011}. 

The forces due to the fluid, as well as the forces due to the dampers
are employed with Newton's second law to yield the body motions. The first two \eqref{eq:mo1}-\eqref{eq:mo2} equate the vertical (heave) forces with the masses and accelerations of the upper and lower cylinder, respectively. The next \eqref{eq:mo3}-\eqref{eq:mo4} are those for the horizontal (sway) forces. The final pair \eqref{eq:mo5}-\eqref{eq:mo6} equate the torques about the $y-$axis to the angular acceleration times moment of inertia of the upper and lower cylinder, respectively.
\begin{align}
  &F_{z1}+F_{z1\rightarrow z1}+F_{z2\rightarrow z1} +F_{hs,z1}+F_{d,z1} =-\omega^2\zeta_{z10}M_1, \label{eq:mo1}\\
  &F_{z2}+F_{z1\rightarrow z2}+F_{z2\rightarrow z2} +F_{hs,z2}+F_{d,z2} =-\omega^2\zeta_{z20}M_2, \label{eq:mo2}\\
  &F_{x1}+F_{x1\rightarrow x1}+F_{x2\rightarrow x1}+F_{y1\rightarrow x1}+F_{y2\rightarrow x1} =-\omega^2\zeta_{x10}M_1, \label{eq:mo3}\\
  &F_{x2}+F_{x1\rightarrow x2}+F_{x2\rightarrow x2}+F_{y1\rightarrow x2}+F_{y2\rightarrow x2} =-\omega^2\zeta_{x20}M_2, \label{eq:mo4}\\
  &F_{y1}+F_{x1\rightarrow y1}+F_{x2\rightarrow y1}+F_{y1\rightarrow y1}+F_{y2\rightarrow y1} +F_{hs,y1}+F_{d,y1} =-\omega^2\theta_{10}I_1, \label{eq:mo5}\\
  &F_{y2}+F_{x1\rightarrow y2}+F_{x2\rightarrow y2}+F_{y1\rightarrow y2}+F_{y2\rightarrow y2} +F_{hs,y2}+F_{d,y2} =-\omega^2\theta_{20}I_2, \label{eq:mo6}
\end{align}
The terms appearing in the above equations have the following interpretation:
\vspace{1em}
\begin{table}[h!]
\renewcommand{\arraystretch}{1.3}
\begin{tabular}{@{} l m{9cm} @{}}
$F_{xi}, \, F_{yi}, \, F_{zi}, \, i \in \{1,2\}$  &  {Exciting forces/torques on cylinder $i$ in the $x,y,$ and $z$ directions}  \\
$F_{\alpha i \rightarrow \beta j}, i,j \in \{1,2\}, \, \alpha,\beta \in \{x,y,z\}$  &  Hydrodynamic force/torque of the $\alpha$ motion of cylinder $i$  in the $\beta$ direction of cylinder $j.$ \\
$F_{hs,yi}, \, F_{hs,zi}, \, i \in \{1,2\}$  &  Hydrostatic forces in the $y$ and $z$ direction on cylinder $i.$ \\
$F_{d,yi}, \, F_{d,zi}, \, i \in \{1,2\}$  &  Damping forces in the $y$ and $z$ direction on cylinder $i.$  \\
$\zeta_{xi0}, \, \zeta_{zi0}, \, \theta_{i0}, \, i \in \{1,2\}$  & Displacement amplitudes of cylinder $i$ in sway ($\zeta_x$), heave ($\zeta_z$), and roll ($\theta$). \\
$M_i, \, i \in \{1,2\}$  & Mass of cylinder $i.$ \\
$I_i, \, i \in \{1,2\}$  & Moment of inertia of cylinder $i$ about the $y$-axis.
\end{tabular}
\end{table}
\vspace{1em}

\noindent The masses and moments of inertia have the explicit form (see \eqref{eq: cylinder sizes}, \eqref{casestudy})
\begin{align*}
& M_1=M_2 = \rho\pi q^3, \\
& I_1=\frac{73}{108} \rho\pi q^5 , \\
& I_2=\frac{757}{108} \rho\pi q^5 ,
\end{align*}
and the damping forces are
\begin{align*}
&F_{d,z1}= - i\omega C(\zeta_{z10}-\zeta_{z20})e^{i\omega t}, \\
&F_{d,z2}=  i\omega C(\zeta_{z10}-\zeta_{z20})e^{i\omega t}, \\
&F_{d,y1}=- \frac{1}{2}i\omega C R^2(\theta_{10}-\theta_{20})e^{i\omega t}, \\
&F_{d,y2}= \frac{1}{2}i\omega C R^2(\theta_{10}-\theta_{20})e^{i\omega t}.
\end{align*}
After the displacements of the cylinders are obtained, the captured power can then be calculated as follows:
\begin{equation}
P_a=\frac{1}{2} C \omega^2 (\zeta_{z10}-\zeta_{z20})(\zeta_{z10}^{*}-\zeta_{z20}^{*})
+\frac{1}{4} C \omega^2 R^2 (\theta_{10}-\theta_{20})(\theta_{10}^{*}-\theta_{20}^{*}),\label{Pa}
\end{equation}
where $\zeta_{zj0}^{*}$ and $\theta_{j0}^{*}$ are the complex conjugates of $\zeta_{zj0}$ and $\theta_{j0}$, respectively.

\section{Design of the WEC for monochromatic waves}
\label{sec:design of the wec}

We now undertake to examine the design of the WEC, based on the three parameters characterizing the environmental conditions $\rho, \, g,$ and $U,$ the two WEC parameters $q$ and $C,$ and the seven WEC performance parameters calculated from the wave-structure interaction problem $P_a, \, \zeta_{x10}, \, \zeta_{x20}, \, \zeta_{z10}, \, \zeta_{z20}, \, \theta_{10}$ and $\theta_{20}.$  An application of Buckingham's $\pi$ theorem \cite{Crowe2001} yields that there will be nine dimensionless quantities that characterize this problem: $\frac{q}{U^2/g}$, $\frac{C}{\rho U^5/g^2}$, $\frac{P_a}{\rho U^7/g^2}$, $\frac{\zeta_{zj0}}{U^2/g}$, $\frac{\zeta_{xj0}}{U^2/g}$, and $\theta_{j0}.$ In the sequel, we will make use of a $\sim$ to denote nondimensional variables, i.e., the nine dimensionless quantitites above will be $\tilde{q}, \, \tilde{C}, \tilde{P_a}, \tilde{\zeta}_{zj0}, \tilde{\zeta}_{xj0},$ and $\tilde{\theta}_{j0}.$

\subsection{The WEC in heave motion under a monochromatic wave}
\label{subsec:heave only}
For simplicity of presentation and ease of understanding we initially consider only the heave mode, motivated by the fact that, while sway and roll are generally coupled, they are both independent of heave. The response of the WEC under incoming monochromatic waves is first considered, where our physical test-case corresponds to a monochromatic wave of wavelength $96.3$ m equal to that at the peak of the design spectrum $S_d,$ and an amplitude $a_d = 0.87 \text{ m,}$ such that the total energy density of the wave is equal to that of $S_d,$ see \eqref{eq:monochrom ampl} and Table \ref{SwSs}.

\subsubsection{Step 1: determination of the WEC's size}
\begin{figure}[H]
    \centerline
    {\includegraphics[scale=0.35,angle=270]{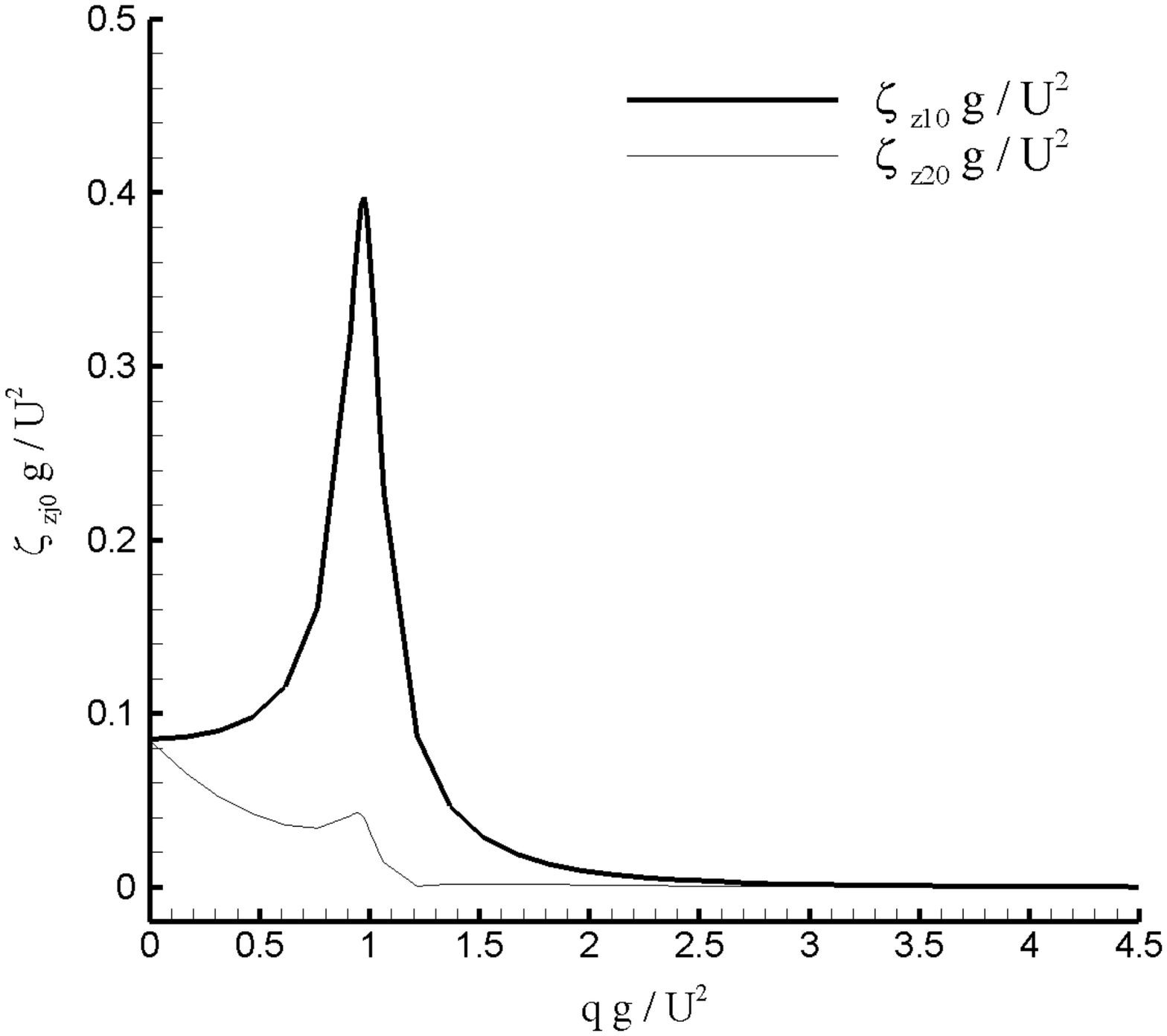}}
    \caption{Displacement amplitudes for each of the two freely floating cylinders ($C=0$) under the design monochromatic wave. $\tilde{\zeta}_{z10}$: upper cylinder, thick line; $\tilde{\zeta}_{z20}$: lower cylinder, thin line.}
  \label{fig:free-heave-z}
\end{figure}

We first choose the dashpot coefficient $C$ to be zero, which means that the two cylinders are freely floating.
In this case, once the incident monochromatic wave is given, the only WEC parameter to be determined is $q$. Dimensional analysis can then be applied to the problem of determining the quantity of interest $q$ for the motions of the upper cylinder $\zeta_{z10}$ and the lower cylinder $\zeta_{z20}$ separately. Once again, Buckingham's $\pi$ theorem yields that, for the variables $\rho,g, U,q,$ and $\zeta_{zj0},$ there exist exactly two nondimensional quantities, which must be related by a relation
\begin{equation} \label{eq:heave-only disp max}
\tilde{\zeta}_{zj0}=\Psi_{1j}(\tilde{q}).
\end{equation}
The maximum displacement of the cylinder $j$ as a function of size $\tilde{q}$ thus corresponds to the extrema of $\Psi_1.$ 
Equation \eqref{eq:heave-only disp max} is plotted in Figure \ref{fig:free-heave-z} for the upper and lower cylinders.
As we are ultimately interested in relative displacements of the cylinders, the global maximum of $\Psi_1(\tilde{\zeta}_{z10})$ and the local maximum of $\Psi_1(\tilde{\zeta}_{z20} )$ which occur at $\tilde{q}=0.97$ yield the chosen design size.
\subsubsection{Step 2: determination of the dashpot coefficient $C$}
The maximum displacement in Fig.\ \ref{fig:free-heave-z} is related to the resonance between the cylinders and the incident monochromatic wave. The introduction of a damper, while changing the magnitude of the displacement, can be shown to have no effect on the location of the resonant maximum, which remains $\tilde{q}=0.97$ (see Fig.\ \ref{fig:free-heave-z}) even for various values of $C$. Thus, the size of the WEC determined from the freely floating case is used to specify the dashpot coefficient $C$.

\begin{figure}[H]
    \centerline
    {\includegraphics[scale=0.35,angle=270]{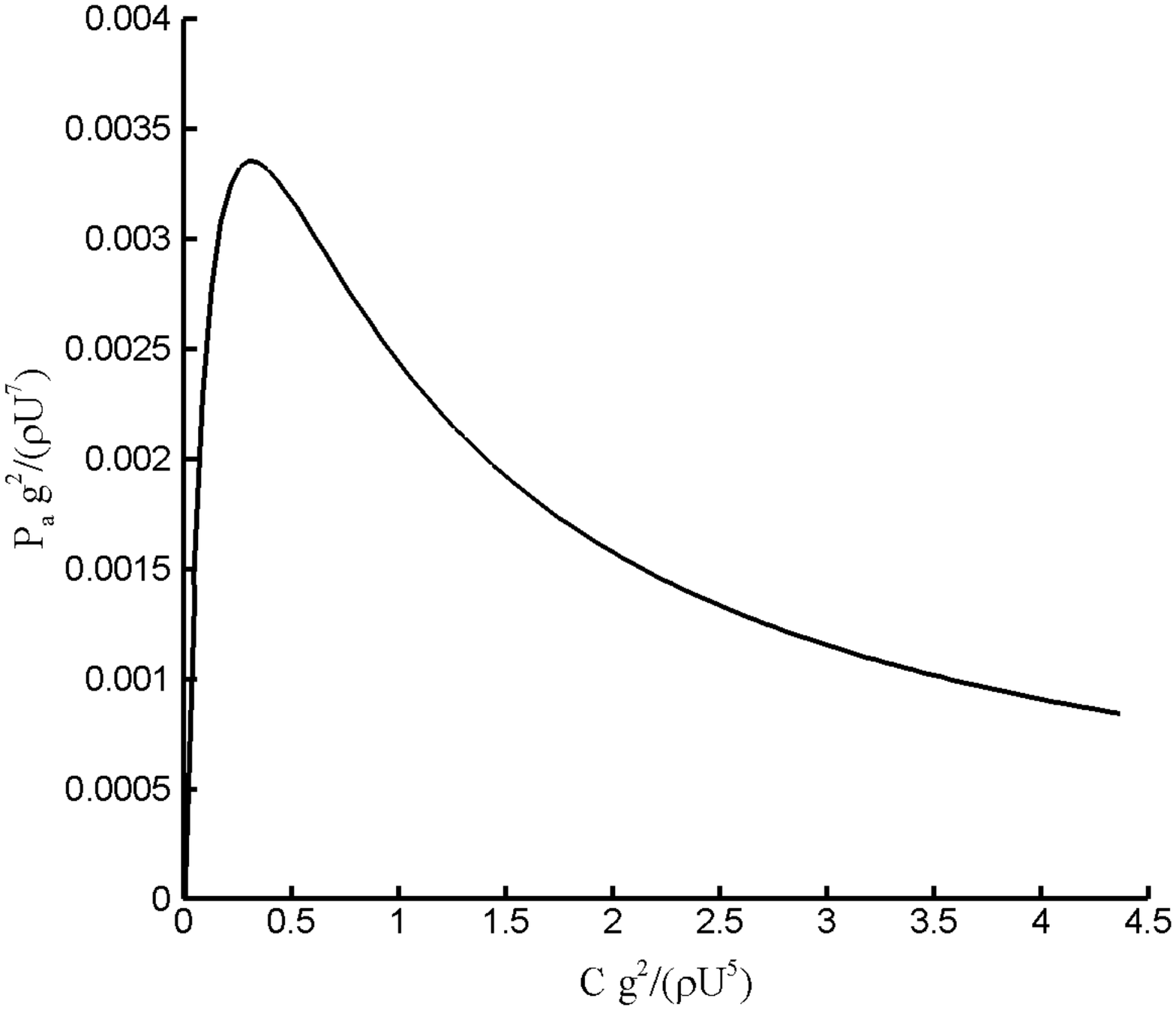}}
    \caption{The relationship between the captured power $\tilde{P}_a$ and the dashpot coefficient $\tilde{C}$ for the heave motion induced by the design monochromatic wave, where $\tilde{q}=0.97$.}
  \label{fig:pald-heave}
\end{figure}

Given the unique relationship between $q$ and $\zeta_{zj0},$ independent of $C,$ described above, the dimensional analysis of the power absorption involves the quantities $\rho, g, U, C, q$ and $P_a,$ which can be written in three dimensionless ratios:
\begin{equation}
\tilde{P}_a=\Psi_2(\tilde{C};\tilde{q}),
\end{equation}
where $\Psi_2$ is plotted in Fig.\ \ref{fig:pald-heave} for the WEC size as determined in the last section ($\tilde{q}=0.97$).

We elect to determine the dashpot coefficient $C$ from the maximum of captured power $P_a$ in Figure \ref{fig:pald-heave}, calculated from the heave terms only in (\ref{Pa}). This results in $ \tilde{C}=0.32$ and $ \tilde{P}_a=0.0034$.

Thus, the WEC design for a monochromatic wave has been determined. Taking the design wave introduced in the beginning of Section \ref{sec:design of the wec} as a physical example, the WEC has the dimensions  $q=9.9 m$ and $C=3.3\times10^5 Ns/m$, and can capture $P_a=3.5\times10^5 \text{ Watt}$ from a monochromatic wave 96.3 m long and 0.87 m in amplitude.

\subsection{General motions of the WEC in monochromatic waves}

Having treated the simpler case of heave-only motion, we now consider the general case in which the WEC is additionally allowed to undergo sway and roll motions. Akin to the previous section which only dealt with the heave motion, the design procedure of the WEC in the general case is also divided into two steps, as illustrated in detail below.

\subsubsection{Step 1: Determination of the WEC's size $q$}
We start with the freely floating case, where the dashpot coefficient $C=0$. Using the equations of motion (\ref{eq:mo1})-(\ref{eq:mo6}), we can obtain the displacements of the two cylinders in the $x$ and $z$ directions, and the angle around the $y$ axis.

\begin{figure}[H]
  \centering
  \subfigure[upper cylinder]{
    \label{fig:freefloat-3m:a} %% label for first subfigure
    \includegraphics[width=2.5in]{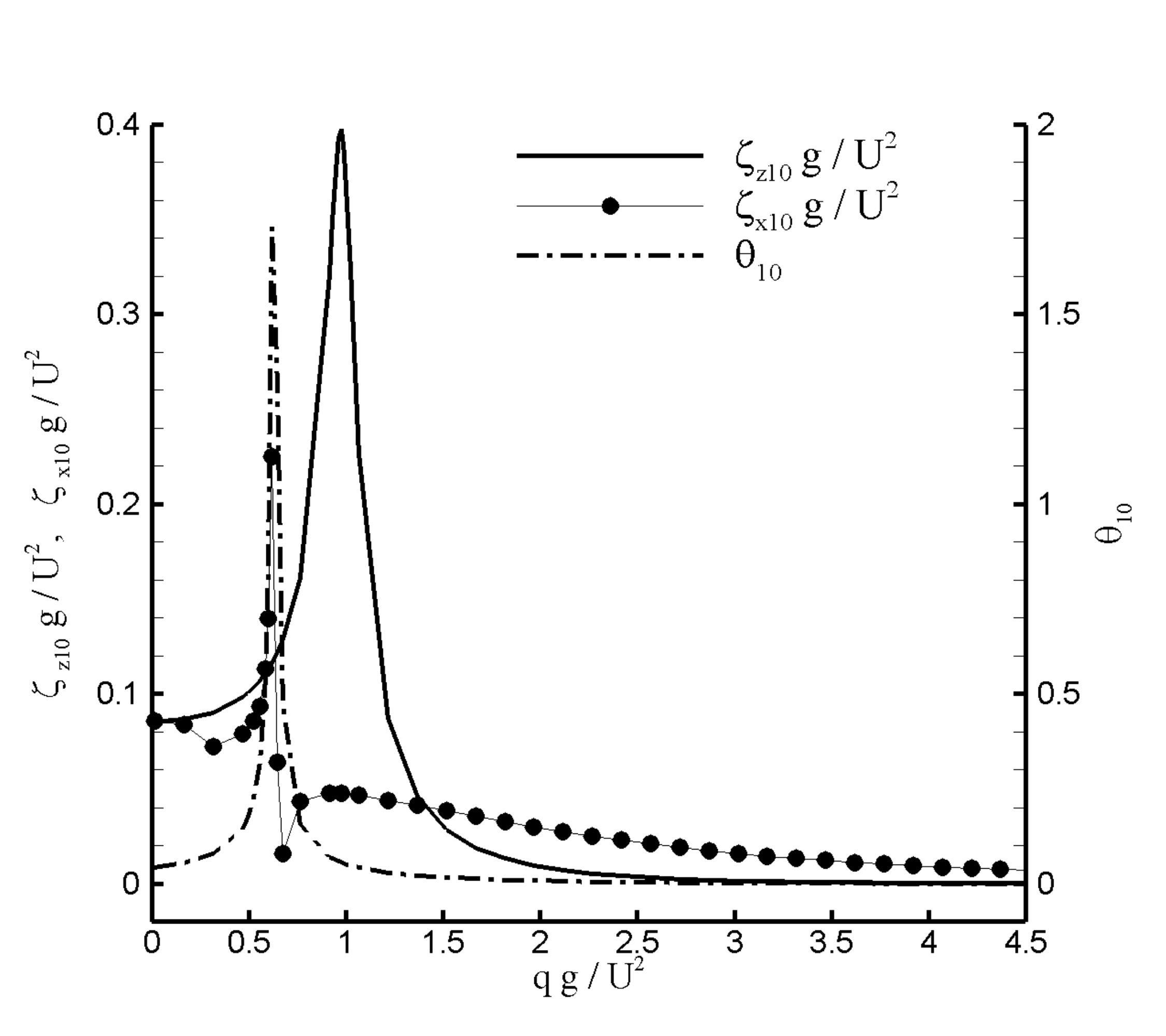}}
  %\hspace{1in}
  \subfigure[lower cylinder]{
    \label{fig:freefloat-3m:b} %% label for second subfigure
    \includegraphics[width=2.5in]{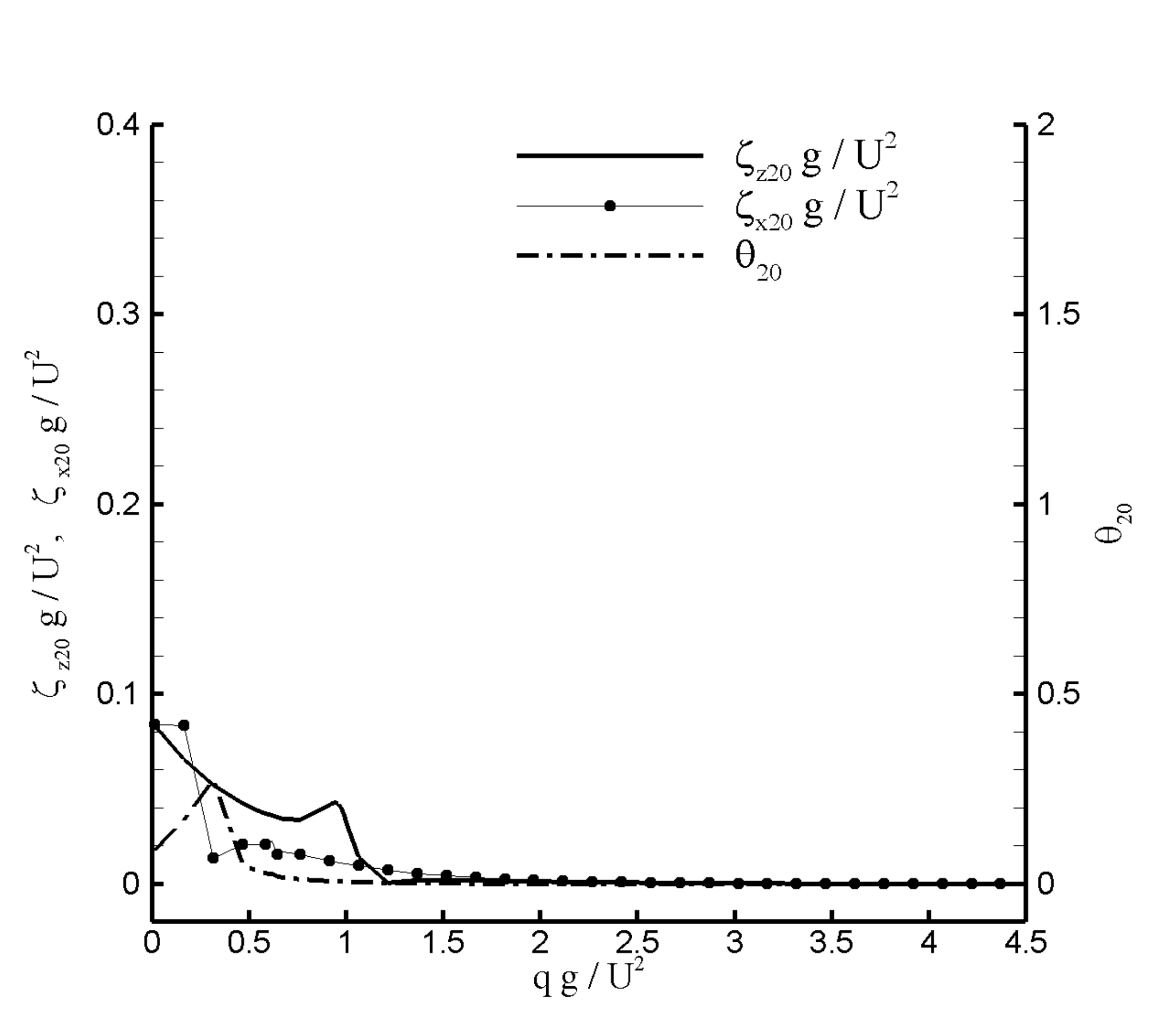}}
  \caption{Displacement amplitudes of the two freely-floating cylinders ($C=0$) in heave, sway and roll under the design monochromatic wave. $\tilde{\zeta}_{zj0}$: amplitude of the vertical displacement; $\tilde{\zeta}_{xj0}$: amplitude of the horizontal displacement; $\theta_{j0}$: amplitude of the angle around the $y$ axis. $j=1,2$ correspond the the upper and lower cylinders, respectively. }
  \label{freefloat-3m} %% label for entire figure
\end{figure}

Once the monochromatic wave is given, or equivalently, once the mean wind speed for the corresponding PM spectrum is given, the physical process of determining the size of the WEC can be written in the following dimensionless form:
\begin{eqnarray}
\tilde{\zeta}_{zj0}&=&\Psi_{1j}(\tilde{q}),\\
\tilde{\zeta}_{xj0}&=&\Psi_{2j}(\tilde{q}),\\
\theta_{j0}&=&\Psi_{3j}(\tilde{q}),\\
\end{eqnarray}
where $\zeta_{zj0}$ and $\zeta_{xj0}$ denote the amplitudes of the vertical and horizontal displacements respectively, $\theta_{j0}$ is the amplitude of the angle around the $y$ axis, and  $j=1,2$ corresponds to the upper and lower cylinder, respectively.
We now seek the maxima of the functions $\Psi_{1j}, \Psi_{2j}$ and $\Psi_{3j},$ presented in Fig.\ref{freefloat-3m}. 

Due to the increase in number of modes, the picture of the displacements is more complex than in the preceding section. It may be observed that the heave mode is decoupled from the sway and roll modes, yielding again the global maximum at $\tilde{q}=0.97.$ The sway and roll modes are coupled, and are observed to present a global maximum for relative displacement at $\tilde{q}=0.61,$ resulting in an ambiguous situation for determining the size of the WEC.

\subsubsection{Step 2: Determination of the dashpot coefficient $C$}
\begin{figure}
    \centerline
    {\includegraphics[scale=0.35,angle=270]{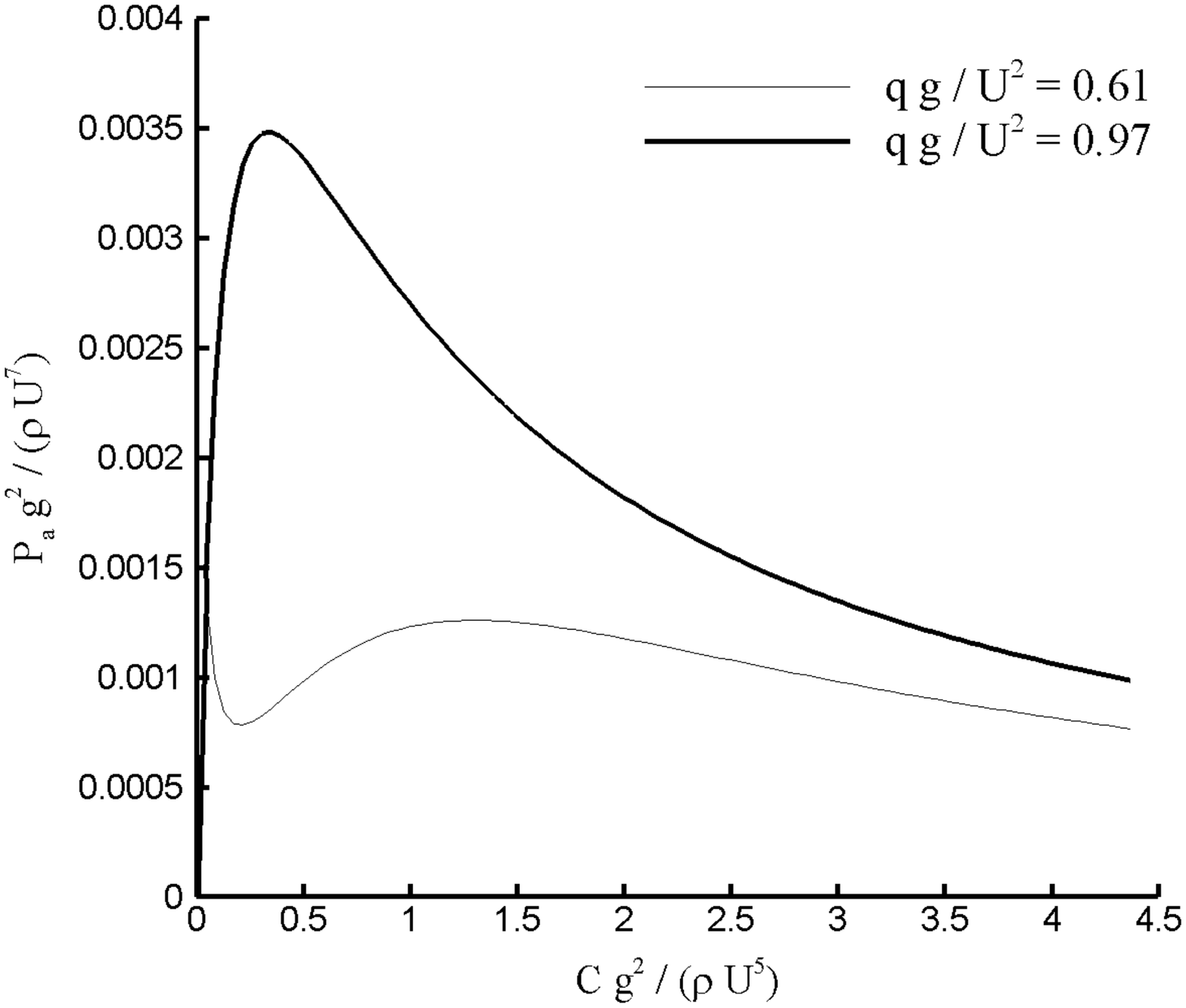}}
    \caption{The captured power $P_a$ of the WEC in the combined motions versus the dashpot coefficient $C$. Thin line: $\frac{q}{U^2/g}=0.61$; thick line: $\frac{q}{U^2/g}=0.97$.}
  \label{fig:pald-3m}
\end{figure}

As in the preceding section, we now suppose that the size of the WEC is given. The captured power $P_a$ then depends on the dashpot coefficient $C$. The determination of optimal power absorption as a function of dashpot coefficient is described in dimensionless form by
\begin{equation}
\tilde{P_a}=\Psi_4(\tilde{C};\tilde{q}),
\end{equation}
where, as we have seen, there is some flexibility in choice of $q.$ The function $\Psi_4$ is plotted in Fig. \ref{fig:pald-3m} for both $\frac{q}{U^2/g}=0.61$ and $\frac{q}{U^2/g}=0.97$. For the device operating optimally in heave ($qg/U^2 = 0.97,$ thick line) there is a unique maximum at $\tilde{C} = 0.34$ with $\tilde{P_a}=0.0035$ (denoted Case E), very close to the heave-only case discussed in Section \ref{subsec:heave only}. For the roll--sway optimized device ($qg/U^2 = 0.61,$ thin line) there are two local maxima $\tilde{C}=0.035$ and $\tilde{C}=1.34,$ with corresponding $\tilde{P_a} = 0.0012$ and $0.0013,$ (denoted Case A1 and A2) respectively.

The situation for monochromatic incident waves is summed up in Table \ref{3m:mono:AH}, which shows the nondimensional size, optimal damping, captured power, and displacement amplitudes for the cases discussed above. As we have observed, introducing roll and sway motions leads to a two-fold branching in the design procedure. Firstly, in free motion, one value of $\tilde{q}$ is found to yield the largest roll and sway displacements, while another value yields the largest heave displacements. While the heave-optimized case has a unique maximum $\tilde{P}_a$ as a function of damping, the roll/sway-optimized case admits two local maxima of $\tilde{P}_a,$ one with relatively low, the other with relatively high damping $\tilde{C},$ compared to the heave case (see Figure \ref{fig:pald-3m}). 

This opens up the possibility that the overall optimal design may not coincide with a design optimized for roll/sway or heave alone, but occupying some middle ground.
The performance of such intermediate devices (Cases B, C, and D), as well as devices somewhat larger than Case E (Cases F, G and H) is explored for the monochromatic design wave in Table \ref{3m:mono:AH}. In each of Cases A1 through H, a damping ${C}$ has been chosen to maximize the captured power.

\begin{table}[!htb]
\renewcommand{\arraystretch}{1.3}
\caption{The size, damping, displacement and captured power of 3-mode WECs in monochromatic waves.} % title name of the table
\centering % centering table
\begin{tabular}{@{} llllllllll @{}}
  \toprule
  &A1&A2&B&C&D&E&F&G&H\\
 \midrule
 $\tilde{q}$ & 0.61 & 0.61 &0.70&0.79&0.88&0.97&1.06&1.15&1.24\\
  
$\tilde{C}$& 0.035&1.34 &0.90&0.67&0.51&0.34&0.56&1.09&1.91\\

$\tilde{P}_a$        &0.0012&0.0013&0.0015&0.0021&0.0028&0.0035&0.0028&0.0020&0.0015\\

$\tilde{\zeta}_{z10}$	 &0.11 &0.089&0.093&0.11&0.14&0.19&0.13&0.063&0.039\\
%\hline
$\tilde{\zeta}_{z20}$	 &0.036&0.053&0.033&0.026&0.023&0.021&0.0097&0.0042&0.0031\\
%\hline
$\tilde{\zeta}_{x10}$	 &0.12 &0.063&0.060&0.055&0.050&0.048&0.047&0.040&0.038\\
%\hline
$\tilde{\zeta}_{x20}$	 &0.018&0.016&0.015&0.013&0.012&0.011&0.0093&0.0069&0.0057\\
%\hline
${\theta}_{10}(rad)$	 &0.70 &0.041&0.052&0.069&0.068&0.056&0.042&0.028&0.022\\
%\hline
${\theta}_{20}(rad)$	 &0.028&0.031&0.021&0.014&0.0094&0.0066&0.0051&0.0035&0.0027\\
\bottomrule
  \end{tabular}
\label{3m:mono:AH}
\end{table}

Here we see that a shift in device size from the smaller, primarily rolling/swaying devices, towards larger, primarily heaving devices has a positive impact on captured power, up to device E. Thereafter, an increase in device size leads to a reduction in captured power, as the larger devices operate preferentially at smaller wavenumbers.

\begin{figure}[H]
  \centering
  \subfigure[Case A1, A2, B and C]{
    \label{fig:pa-k-AH-3m:a} %% label for first subfigure
    \includegraphics[width=2in,angle=270]{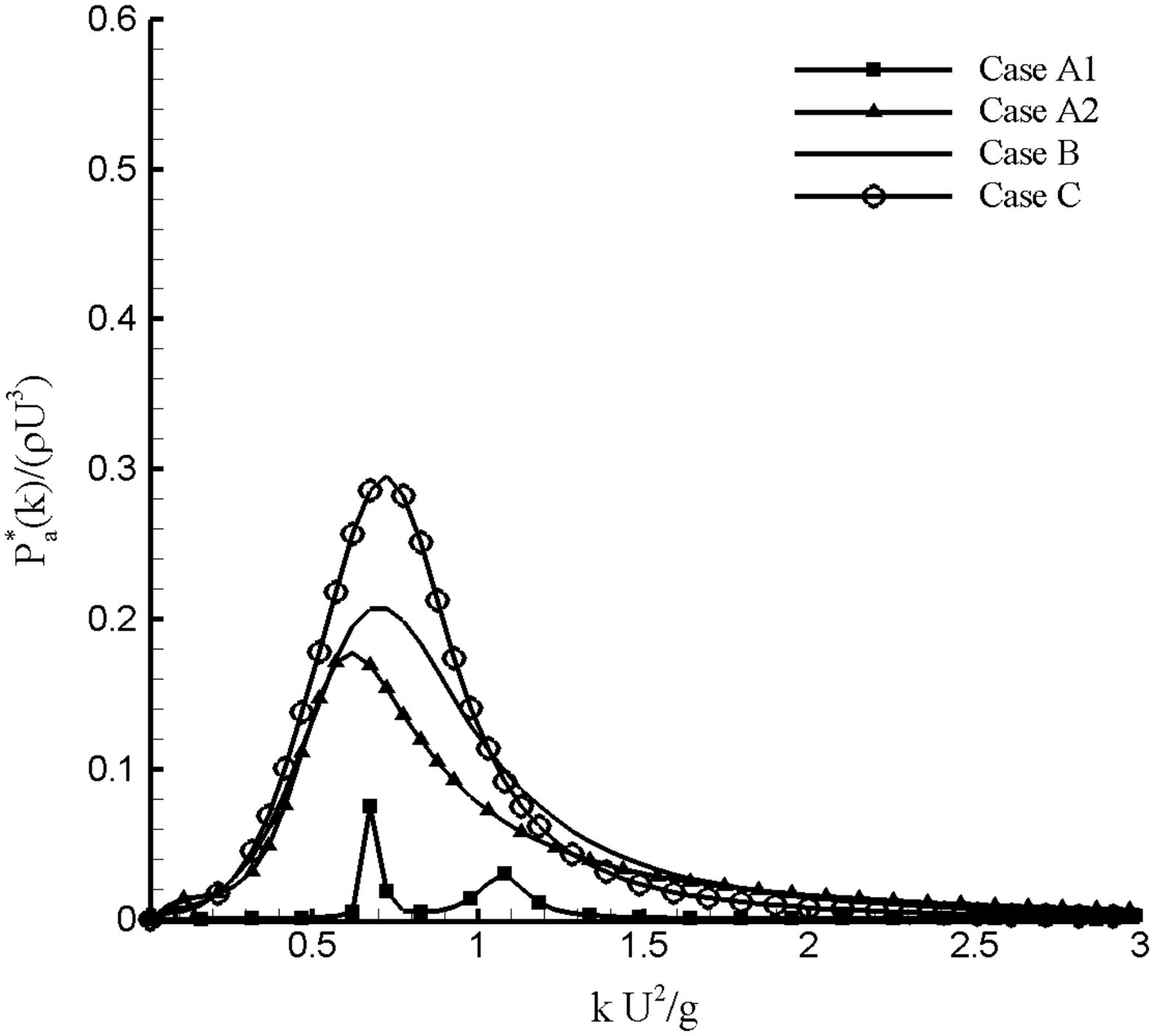}}
  %\hspace{1in}
  \subfigure[Case D and E]{
    \label{fig:pa-k-AH-3m:b} %% label for second subfigure
    \includegraphics[width=2in,angle=270]{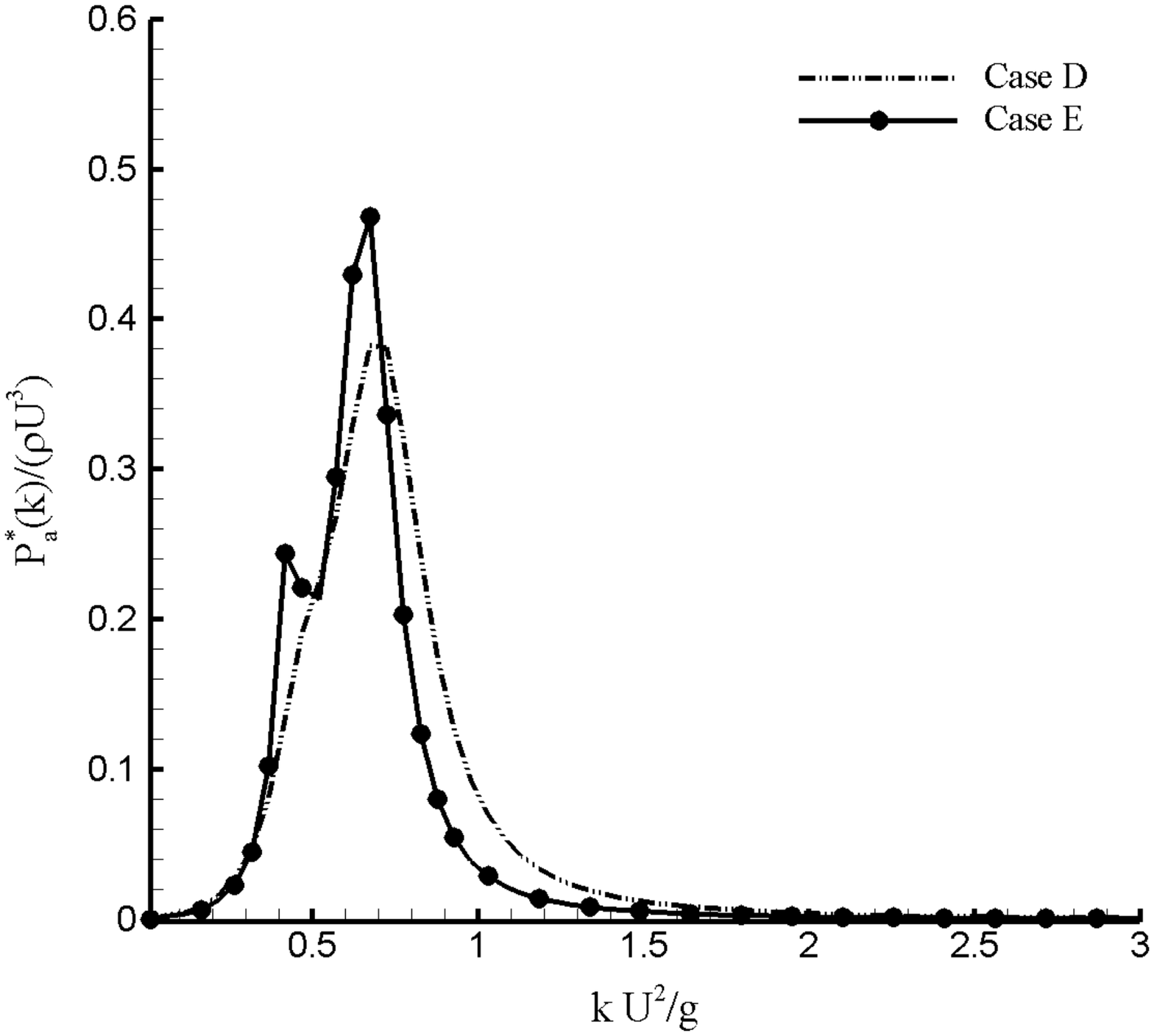}}
  \subfigure[Case F, G and H]{
    \label{fig:pa-k-AH-3m:c} %% label for second subfigure
    \includegraphics[width=2in,angle=270]{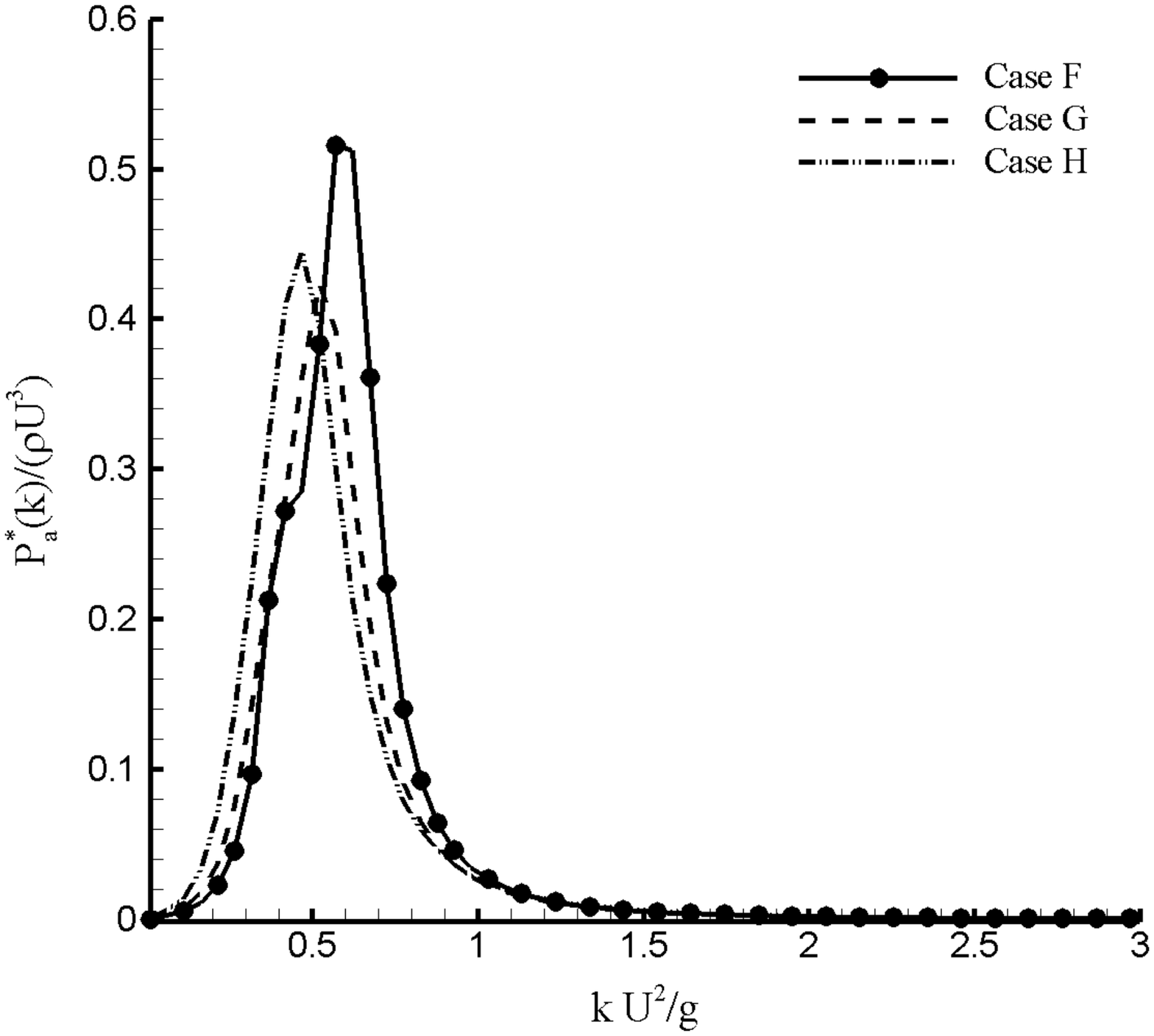}}
  \caption{The dimensionless captured power per unit wave amplitude square $P_a^{*}(k)/(\rho U^3)$ under different monochromatic waves as a function of wavenumber.}
  \label{fig:pa-k-AH-3m} %% label for entire figure
\end{figure}

\begin{figure}[H]
  \centering
  \subfigure[]{
    \label{fig:dis1-k-AH-3m:a} %% label for first subfigure
    \includegraphics[width=2.1in,angle=270]{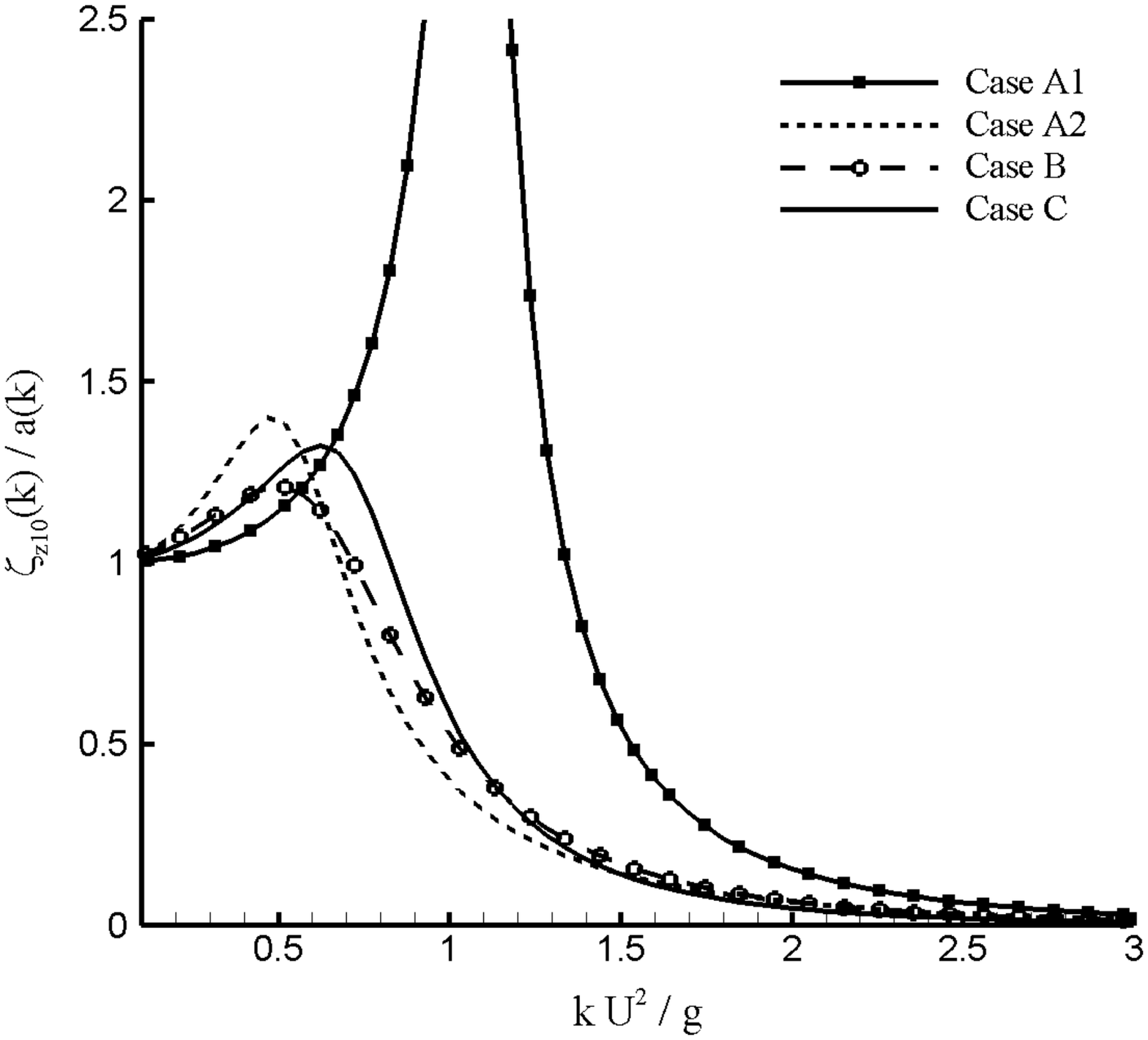}}
  %\hspace{1in}
  \subfigure[]{
    \label{fig:dis1-k-AH-3m:b} %% label for second subfigure
    \includegraphics[width=2.1in,angle=270]{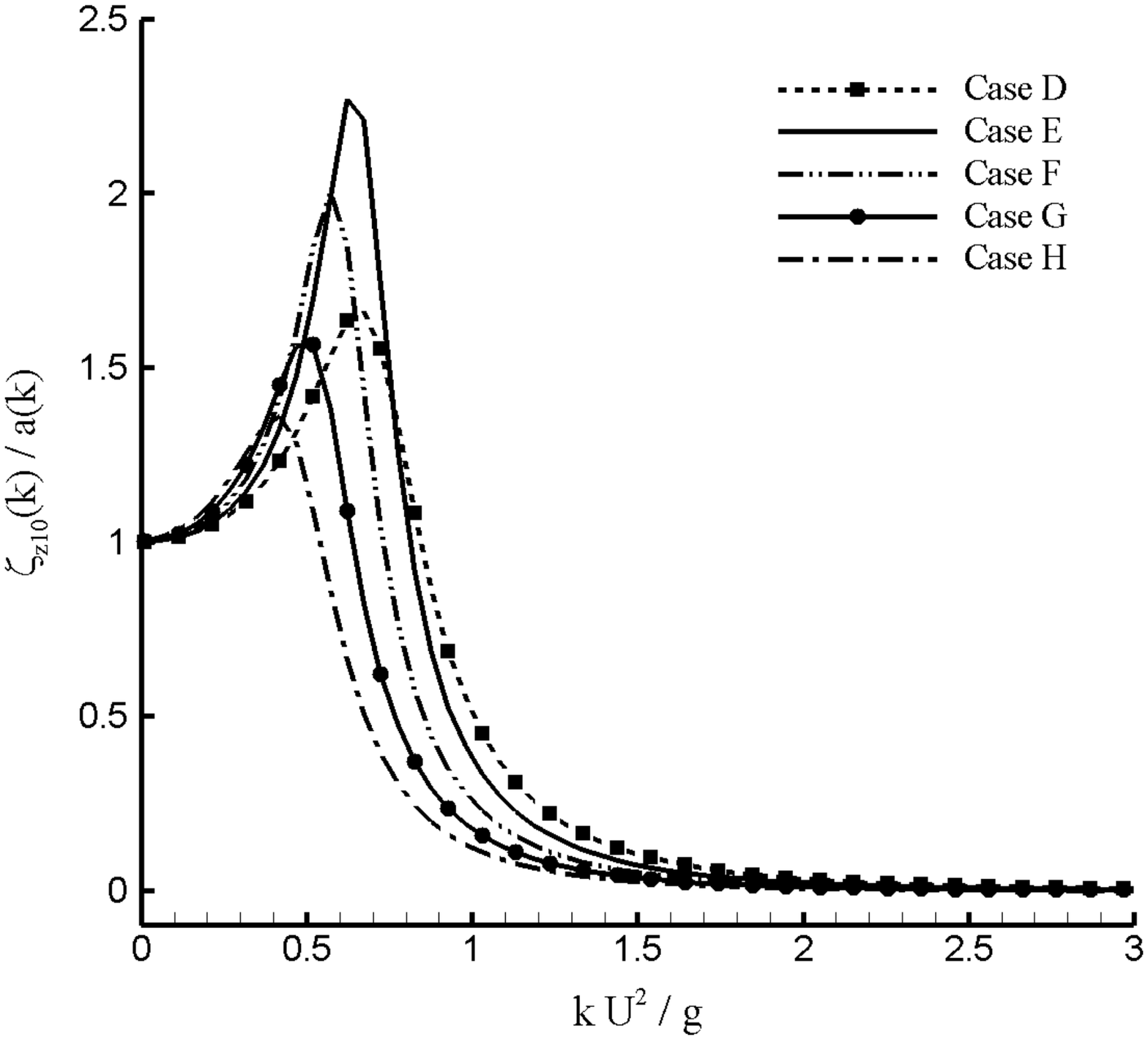}}
  \caption{The dimensionless displacement $\zeta_{z10}/a_0$ of the upper cylinder in the combined motions under different monochromatic waves.(a)vertical displacement of cases A1, A2, B and C; (b)vertical displacement of cases D, E, F, G and H.}
  \label{fig:dis1-k-AH-3m} %% label for entire figure
\end{figure}

This situation is depicted in Figure \ref{fig:pa-k-AH-3m}, which shows $\tilde{P}_a^* \equiv P_a^*/(\rho U^3),$ the dimensionless captured power per unit wave amplitude squared, where $P_a^* \equiv P_a/a_0^2.$  To illustrate the associated displacements, Figure \ref{fig:dis1-k-AH-3m} shows the displacement in heave for the upper cylinder $\zeta_{z10}$  divided by $a_0$. Note that for case A1, the maximum value of $\zeta_{z10}(k)/a(k)$ is 4.4 (not shown).

\section{Design of the WEC for a PM spectrum}
\label{section: irregular design}
Up to this point, we have considered WEC design for monochromatic waves. In brief: a given wind speed $U$ determines the two necessary parameters, wavenumber $k_p$ and amplitude $a_0$ in terms of the PM spectrum. With a monochromatic wave fully described by $(k_p,a_0),$ we may initially assume freely floating cylinders, and choose their size $\tilde{q}$ for maximum displacement in roll and sway (as these modes are coupled), for maximum displacement in heave, or at some intermediate value. In each case, a damping $\tilde{C}$ is chosen to maximize the captured power for this incident design wave, leading to the cases A1 through H above. As demonstrated in Figures \ref{fig:pa-k-AH-3m} and \ref{fig:dis1-k-AH-3m}, the motions and performance of a device designed for a wave $(k_p,a_0(k_p)$ may change considerably for other waves.

For practical reasons, our primary interest must be focused on irregular waves, where we may elect to tune the device to operate optimally at the peak of the spectrum, but must consider its performance for a broad band of incident waves. We begin with some preliminaries regarding the behavior of the WEC in irregular seas.

For a monochromatic wave, the absorbed power $P_a$ (see \eqref{Pa}) and displacements $\zeta_{\alpha j 0}$ are written as
\begin{align*}
& P_a (q,C,k,a_0) \equiv a_0^2 \hat{P}_a(q,C,k), \\
& \zeta_{\alpha j 0}(q,C,k,a_0) \equiv a_0 \hat{A}_{\alpha j}(q,C,k),\\
& \theta_{j 0}(q,C,k,a_0) \equiv  a_0 \hat{A}_{y j}(q,C,k).
\end{align*}
where $j =1,2$ denotes the upper and lower cylinders, $\hat{P}_a$ is the absorbed power per unit wave amplitude square, and $\hat{A}_{\alpha j}$ with $\alpha \in \{x,y,z\}$ denote the relative amplitudes of sway, roll and heave motions, respectively.
For a given spectrum $S(k)$ the total absorbed power by a device of type $(q,C)$ is 
\begin{equation}
P^{\text{total}}_a = \int_0^\infty 2 \hat{P}_a(k) S(k) dk.
\end{equation}

Just as the spectrum describes the distribution of wave energy among different frequencies, and allows for statistical inferences such as a definition of the significant wave-height, so analogously we may consider a \emph{displacement spectrum}
\begin{equation}
E_{\alpha j}(k) \equiv S(k) (\hat{A}_{\alpha j})^2,
\end{equation}
and define the \emph{significant displacement} by 
\begin{equation}
H_{\alpha j}^{1/3} = 4 \cdot \left(\int_0^\infty E_{\alpha j}(k) dk\right)^{1/2}.
\end{equation}
Here $H_{\alpha}^{(1/3)}$ ($\alpha=x,y,z$) is the the distance from the displacement's trough to crest and
\begin{eqnarray}
\zeta_{zj0}^{(1/3)}=\frac{1}{2}H_{zj}^{(1/3)}, \label{a1} \\
\zeta_{xj0}^{(1/3)}=\frac{1}{2}H_{xj}^{(1/3)},\label{a2}  \\
\theta_{j0}^{(1/3)}=\frac{1}{2}H_{yj}^{(1/3)},\label{a3}
\end{eqnarray}
are the so-called ``significant amplitudes of the displacement" in $z$ and $x$ directions, and the angle around $y$ axis, respectively.

\begin{table}[!ht]
\renewcommand{\arraystretch}{1.3}
\caption{The nondimensional significant amplitudes of displacement in heave $\tilde{\zeta}^{(1/3)}_{zj0},$ sway $\tilde{\zeta}^{(1/3)}_{xj0},$ and roll $\theta_{j0}^{(1/3)}$ (rad), along with dimensionless captured power $\tilde{P}_a^{\text{total}}$ for the WECs A1 through H attacked by a design spectrum $S_d,$ a severe spectrum $S_{s1}$ ($U_{s1} = 1.5 U_d$), and a second severe spectrum $S_{s2}$ ($U_{s2} = 2 U_d$).} % title name of the table
\centering % centering table
\begin{tabular}{@{} llllllllll @{}}
\toprule
  &A1&A2&B&C&D&E&F&G&H\\
  \midrule
 $\tilde{q}$ &0.61 & 0.61 &0.70&0.79&0.88&0.97&1.06&1.15&1.24\\
  
$\tilde{C}$& 0.035&1.34 &0.90&0.67&0.51&0.34&0.56&1.09&1.91\\
\midrule
  $S_{d}$\\

$\tilde{P}_a^{\text{total}}$ & $7.01\times10^{-5}$&0.00065&0.00081&0.0010&0.0011&0.0010&0.0011&0.00089&0.00086\\

$\tilde{\zeta}_{z10}^{(1/3)}$	 &0.22&0.097&0.096&0.11&0.12&0.14&0.12&0.094&0.071\\

$\tilde{\zeta}_{z20}^{(1/3)}$	 &0.044&0.065&0.040&0.030&0.027&0.025&0.021&0.017&0.013\\

$\tilde{\zeta}_{x10}^{(1/3)}$	 &0.10&0.077&0.073&0.068&0.064&0.059&0.056&0.054&0.051\\

$\tilde{\zeta}_{x20}^{(1/3)}$	 &0.022&0.019&0.018&0.016&0.015&0.014&0.012&0.011&0.0093\\

${\theta}_{10}^{(1/3)}$	 &0.60&0.050&0.059&0.073&0.084&0.096&0.068&0.045&0.032\\

${\theta}_{20}^{(1/3)}$	 &0.012&0.015&0.0093&0.0058&0.0039&0.0027&0.0020&0.0016&0.0012\\
\midrule
  $S_{s1}$ \\

$\tilde{P}_a^{\text{total}}$ & 0.00016&0.0025&0.0030&0.0037&0.0043&0.0046&0.0061&0.0060&0.0074\\

$\tilde{\zeta}_{z10}^{(1/3)}$	 &0.37&0.31&0.29&0.29&0.32&0.35&0.34&0.31&0.29\\

$\tilde{\zeta}_{z20}^{(1/3)}$	 &0.15&0.25&0.18&0.14&0.13&0.12&0.11&0.11&0.10\\

$\tilde{\zeta}_{x10}^{(1/3)}$	 &0.25&0.21&0.22&0.21&0.21&0.21&0.20&0.19&0.19\\

$\tilde{\zeta}_{x20}^{(1/3)}$	 &0.090&0.076&0.081&0.081&0.078&0.075&0.070&0.065&0.060\\

${\theta}_{10}^{(1/3)}$	 &0.98&0.16&0.14&0.17&0.21&0.29&0.23&0.16&0.12\\

${\theta}_{20}^{(1/3)}$	 &0.13&0.17&0.11&0.075&0.058&0.046&0.039&0.034&0.029\\
\midrule
  $S_{s2}$ \\

$\tilde{P}_a^{\text{total}} $ & 0.00019&0.0044&0.0049&0.0056&0.0064&0.0068&0.0097&0.011&0.014\\

$\tilde{\zeta}_{z10}^{(1/3)}$	 &0.55&0.54&0.51&0.52&0.53&0.56&0.56&0.54&0.53\\

$\tilde{\zeta}_{z20}^{(1/3)}$	 &0.34&0.47&0.39&0.33&0.30&0.29&0.28&0.27&0.27\\

$\tilde{\zeta}_{x10}^{(1/3)}$	 &0.45&0.42&0.42&0.42&0.42&0.42&0.42&0.41&0.40\\

$\tilde{\zeta}_{x20}^{(1/3)}$	 &0.20&0.19&0.19&0.20&0.20&0.19&0.19&0.18&0.17\\

${\theta}_{10}^{(1/3)}$	 &1.07&0.317&0.21&0.22&0.27&0.37&0.31&0.24&0.19\\

${\theta}_{20}^{(1/3)}$	 &0.31&0.40&0.25&0.18&0.14&0.12&0.10&0.088&0.077\\
\bottomrule
  \end{tabular}
\label{3m:severe2:AH}
\end{table}

Applying the concepts developed above to the problem of power absorption from an incident, broad-banded sea, we evaluate the above expressions for the spectra introduced in Section \ref{sec:physical preliminaries}. The results are given in nondimensional form in Table \ref{3m:severe2:AH}, which shows the captured power and displacement amplitudes for the spectra $S_d, \, S_{s1}$ and $S_{s2},$ nondimensionalized by $U=U_d.$ This may be compared to the analogous Table \ref{3m:mono:AH} for the monochromatic case. In the following section, we turn to a discussion of these results.

\section{Discussion}
\label{section:discussion}
\noindent As we have mentioned above, several competing criteria exist in determining WEC size. Those we shall consider in depth are limited to power capture, which naturally should be maximized, and survivability -- measured in terms of device motions. 

We note that the Cases A1 through H presented above are ordered by increasing size $q$ which may be assumed correlated to the cost \emph{per device,} all other things being equal. Due to the burgeoning state of wave energy technology, it seems premature to speculate any further about cost, given that it depends not only on device size, but also design specifics such as materials and component costs, as well as costs related to regular maintenance or major overhaul, both factors which will in turn be affected by size. 

In the following sections, we will delve into a detailed analysis of the WEC behaviour with a view to power capture and survivability. Subsequently, a synthesis of these two viewpoints is attempted, bearing in mind the primary aim of providing quantitative information relating to the design of oscillating body converters in a range of different, broad-banded sea states.

\subsection{Power capture}
\label{subsec:power capture}
The most straightforward metric to evaluate concerns the power captured by a WEC. For a design PM spectrum $S_d$ corresponding to a wind speed $U_d=10$ m/s, and severe spectra $S_{s1}$ and $S_{s2}$ corresponding to $U_{s1} = 15$ and $U_{s2} = 20$ m/s, respectively, the dimensional size, damping and absorbed power of WECS A1 through H are presented in Table \ref{table:dimensional power capture}.

\begin{table}[h]
\centering
\renewcommand{\arraystretch}{1.3}
\caption{Dimensional absorbed power $P_a$ (Watt) for cases A1 through H, for an incoming monochromatic wave ($P_a^m$) and the design PM spectrum with $U = 10$ m/s ($P_a^d$), both with the same energy density of $3.7 \text{ KJ}/\text{m}^2.$ Also given are the absorbed power for the severe spectra $S_{s1}$ ($P_a^{s1}$) and $S_{s2}$ ($P_a^{s2}$).}
\label{table:dimensional power capture}
\begin{tabular}{@{} llllllllll @{} }
\toprule
 & A1&	A2&	B&	C&	D&	E&	F&	G&	H \\
 \midrule
q [m] &6.2 &	6.2	&7.1&	8.1&	9.0&	9.9&	10.8&	11.4&	12.7 \\
C $(\cdot 10^5$) [Ns/m] &	$0.364$	& $14.0$	& $9.37$	& $6.98$& 	$5.31$& 	$3.54$& 	$5.83$& 	$11.3$	& $19.9$
 \\
$P_a^m$ $(\cdot 10^5$) [W] &	$1.25$	& $1.35$ &	$1.56$ &	$2.19$	& $2.92$ &	$3.64$ &	$2.92$	& $2.08$ &	$1.56$ \\
$P_a^{d}$ $(\cdot 10^5$) [W] & $0.0730$	& $0.672$ &	$0.849$ 	& $1.04$ & $1.14	$ & $1.06$ &	$1.17$	& $0.931$ & $0.893$  \\
\midrule
$P_a^{s1}$ $(\cdot 10^5$) [W] & $0.165$	& $2.65$ &	$3.13$ &	$3.80$ &	$4.43$ &	$4.77$ &	$6.36$ &	$6.27$ &	$7.73$ \\
$P_a^{s2}$ $(\cdot 10^5$) [W]& $0.197$ &	$4.62$	& $5.09$	& $5.88$ &	$6.67$ &	$7.10$ &	$10.1$ &	$11.0$ &	$15.0$ \\ 
\bottomrule
\end{tabular}
\end{table}

We recall the monochromatic wave used for device design, with a wavelength of 96.3 m and an amplitude of 0.87 m, with an energy density of 3.7 KJ/m$^2$ equal to that of the design PM spectrum parametrised by a wind-speed $U = 10$ m/s. The picture which emerges from comparing the absorbed powers in the monochromatic and spectral cases is quite striking. While the narrow-banded response of device A1 (see Figure \ref{fig:dis1-k-AH-3m:a}) yields a performance comparable to slightly larger devices for monochromatic waves, power absorption declines drastically for an incident PM spectrum.

Likewise, though the heave-optimized device E is clearly superior to devices of similar size (D and F) for monochromatic waves, this situation sees a dramatic reversal in the case of incident irregular waves. This shows the potential utility of de-tuning in WEC design, as devices on either side of the heave-optimum outperform it for irregular seas. In particular, it demonstrates the pitfalls of a design based on monochromatic waves.

Dimensional values of captured power are also provided for the two severe spectra, $S_{s1}$ corresponding to a wind speed $U_{s1} = 15$ m/s, or an energy density of 18.7 KJ/m$^2$, as well as $S_{s2},$ corresponding to a wind speed $U_{s2} = 20$ m/s and an energy density of 59.6 KJ/m$^2$. As expected, the larger devices benefit most from this increased wave resource, while a sea composed of increasingly long waves ($\lambda_p$ for $S_{s1}$ is 217 m, and for $S_{s2}$ is 385 m, see Table \ref{SwSs}) begins to saturate the power capture capabilities of the smallest devices. In the following sections on survivability and grading of WECs, we shall explore the feasibility of operating WECs in such large sea states.

\subsection{Survivability}
\label{subsec: survivability}
We come now to the less well-defined of the two concepts with a bearing on the performance of a twin-cylinder WEC: survivability. The disparity between the motions and resulting loads experienced by a WEC in normal operation, and those during severe conditions may be immense. Following Brown et al \cite{brown2010towards} we distinguish between the reliability of a WEC, related to failure during normal operation, and survivability. This latter concept applies to sea states outside of the intended operating conditions -- when the average conditions of the ocean environment exceed the safe operational limits of the device.

While it is clear that WECs must be robust in design, as during a ten-year operational period a converter may expect to see some tens of millions of waves, during particularly severe events, power production will need to be halted in order to avoid damage to the device or loss of station-keeping.

\begin{table}[h]
\renewcommand{\arraystretch}{1.3}
\centering
\caption{Relative heave displacements versus draft $\zeta^r_{z} = (\zeta_{z10}^{(1/3)}-\zeta_{z20}^{(1/3)})/q$, and relative roll displacement $\theta^r = \theta^{(1/3)}_{10}/(\pi/2)$ }
\label{table:survivability}
\begin{tabular}{@{} llllllllll @{} }
\toprule
 & A1&	A2&	B&	C&	D&	E&	F&	G&	H \\
 \midrule
 $S_d$ \\
$\zeta_z^{r}$ & \cellcolor{orange!25}0.29	&\cellcolor{green!25}0.05	&\cellcolor{green!25}0.08&\cellcolor{green!25}	0.10&	\cellcolor{green!25}0.11&\cellcolor{green!25}	0.12&\cellcolor{green!25}	0.09&\cellcolor{green!25}	0.07&\cellcolor{green!25}	0.05 \\
$\theta_{1}$&\cellcolor{red!25}0.38&	\cellcolor{green!25}0.03&	\cellcolor{green!25}0.04	&\cellcolor{green!25}0.05&	\cellcolor{green!25}0.05&	\cellcolor{green!25}0.06&\cellcolor{green!25}	0.04&	\cellcolor{green!25}0.03&	\cellcolor{green!25}0.02 \\
$S_{s1}$ \\
$\zeta_z^r$ & \cellcolor{red!25}0.36&	\cellcolor{green!25}0.10	&\cellcolor{yellow!25}0.16	&\cellcolor{yellow!25}0.19	&\cellcolor{yellow!25}0.22&\cellcolor{yellow!25}	0.24	&\cellcolor{yellow!25}0.22&	\cellcolor{yellow!25} 0.18&\cellcolor{green!25}	0.15 \\
$\theta_{1}$ & \cellcolor{red!25}0.62&	\cellcolor{green!25}0.10&	\cellcolor{green!25}0.09&	\cellcolor{green!25}0.11	&\cellcolor{green!25}0.13&	\cellcolor{green!25}0.18&	\cellcolor{green!25}0.15	&\cellcolor{green!25}0.10&	\cellcolor{green!25}0.08 \\
$S_{s2}$ \\
$\zeta_z^r$& \cellcolor{red!25}0.34	&\cellcolor{green!25}0.11&\cellcolor{yellow!25}	0.17	&\cellcolor{yellow!25}0.24	&\cellcolor{orange!25}0.26&\cellcolor{orange!25}	0.28	&\cellcolor{orange!25}0.26	&\cellcolor{yellow!25}0.24&\cellcolor{yellow!25}	0.21 \\
$\theta_{1}$& \cellcolor{red!25}0.68	&\cellcolor{yellow!25}0.20&\cellcolor{green!25}	0.13	&\cellcolor{green!25}0.14	&\cellcolor{green!25}0.17&	\cellcolor{yellow!25}0.24&	\cellcolor{yellow!25}0.20&\cellcolor{green!25}	0.15&	\cellcolor{green!25}0.12 \\
\bottomrule
\bottomrule
\end{tabular}
\end{table}

We have developed an example framework for survivability for the twin-cylinder WEC in the three spectral sea states considered, which is presented in Table \ref{table:survivability}. The maximum allowed relative vertical travel $\zeta^{(1/3)}_{z10}-\zeta^{(1/3)}_{z20}$ is limited to $q/3,$ while the maximum allowable roll is $30^\circ.$ Those cases which exceed these values are marked red (PDF only). A vertical travel of more than $q/4$ or a roll of more than $22.5^\circ$ is marked orange, while a vertical travel of more than $0.15q$ or a roll more than $13.5^\circ$ is marked yellow (PDF only). Device motions smaller than these are marked green (PDF only). Recall that these nondimensional quantities depend only on the relations $U_{s1} = 1.5\cdot U_d$ and $U_{s2} = 2\cdot U_d$ as specified in Section \ref{sec:physical preliminaries}, and the concomitant changes in significant wave-height and peak wavenumber.

For illustrative purposes, if the design spectrum $S_d$ is generated by a fresh breeze ($U_d = 10 m/s,$ or 5 Beafort, 2.47 m $H^{(1/3)}$), then the first severe state $S_{s1}$ may be thought generated by a high wind (7 Beaufort, 5.5 m $H^{(1/3)}$). The second severe state $S_{s2}$ occurs under conditions between gale and severe gale (8--9 Beaufort, 9.9m $H^{(1/3)}$). These extremely harsh conditions represent an energy density more than 16 times that of the design spectrum, and may be expected to challenge the device design. 

While the increase in significant wave-height between the design spectrum $S_d$ and the severe case $S_{s2}$ may seem dramatic, there is no doubt that such conditions will be encountered within the operational life of a WEC. For example, while deep water conditions for the Eastern Mediterranean off Israel's coasts may see significant wave heights greater than 2 m only 6 \% of the time, and wave heights in summer rarely exceed 1--1.5 m, nevertheless storms with $H^{(1/3)}$ in excess of 5 m occur almost yearly. The 10-year return period significant wave height is nearly 7 m, which clearly falls  within the expected operational life of a converter.

From a pure survivability standpoint, it is immediate only that the smallest converter A1 is not viable. In particular, the very small damping of this configuration (see Table \ref{table:dimensional power capture}), while allowing for efficient power capture from the roll mode, also leads to overly large displacements even for design conditions. With survivability as the central aim of design, larger structures will necessarily fare better, though the differences between devices D, E, and F are in practice rather small. While other authors (e.g.\ Maisondieu \cite{Maisondieu_2015} or Brown et al \cite{brown2010towards}) have investigated survivability of WECs, they have been forced to do so without reference to the hydrodynamics and actual displacements of a floating device, but rather purely based on estimations of the incident wave power.

\subsection{Grading WEC sizes}
\label{subsec:grading}

We shall now make a preliminary attempt to sum up the results of the preceding sections. The intricacies of WEC economics, as well as the many factors which are outside the scope of the present study, such as moorings, specifics of the PTO, control strategies, power conversion and transmission, and other environmental factors from seasonal variability to extreme events, will need to be taken into account for a fuller analysis. In addition, WEC cost will not be considered, and is likely to impact significantly the ultimate design. 

\begin{figure}[h]
\centering
\includegraphics[scale=0.6]{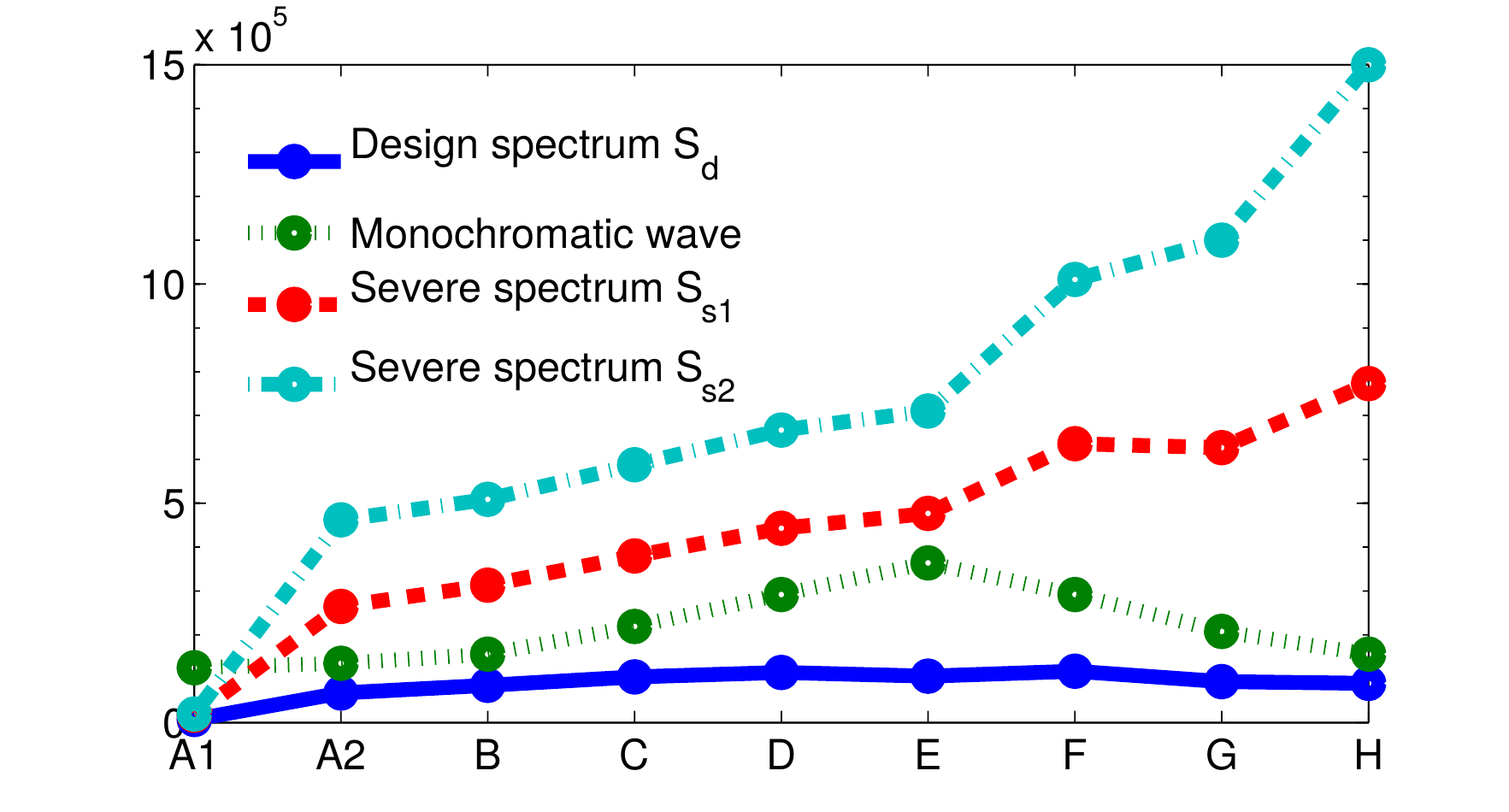}
\caption{Comparison of power absorbed (in Watt) by devices A1 through H under the four conditions -- the design monochromatic wave (dotted line), the design spectrum $S_d$ (solid line), the first severe spectrum $S_{s1}$ (dashed line), and the second severe spectrum $S_{s2}$ (dash-dotted line), where $U_d = 10$ m/s.}
\label{fig: comparison abs power}
\end{figure}

Figure \ref{fig: comparison abs power} compares the dimensional absorbed power for the nine devices considered under different conditions. Once again, we point out that the monochromatic case is essentially artificial, and included only for illustrative purposes. The lessons to be drawn from this comparison will likely change as wave-power technology matures. While current oscillating-body devices may be rather small, and situated in shallow water with the intention of keeping costs down, future developments will naturally lead to a move into the more powerful wave-regimes further offshore (see Stiassnie et al \cite{Stiassnie2015} for a discussion). 

As an example, while there is a 15 \% reduction in absorbed power between Case E and Case H under the design spectrum, the corresponding increase in absorbed power for severe case $S_{s2}$ is upwards of 60 \%. The fact that, off the Eastern Mediterranean Coast, some 45 \% of average wave power comes during storm events that occur only 5 \% of the time indicates the utility of the larger design \cite{Kroszynski1979}.  This is compounded by the increase in potential survivability of the larger devices as indicated in the previous section.
Depending on the variability of the wave-energy resource, more or less weight may ultimately be given to each of the considerations just outlined.  The fact that the larger devices exhibit smaller relative motions may also be a benefit for their reliability, in terms of limiting loading during normal operation. Ultimately, an effort will have to be made to weigh the additional cost of a larger device against the increase in survivability. Both of these in turn will need to be weighed against the potential of continuing operation during high-energy events, while sustaining a slight performance decrease for low-energy sea states.

\section{Conclusions}
\label{sec:conclusions}

We have investigated in detail the hydrodynamics of a model WEC consisting of two floating, axisymmetric cylinders connected at their upper and lower perimeters by a continuously distributed damper -- allowing power capture from heave and roll modes. While other authors have studied various aspects of the problem of floating cylinders, the present work has addresses for the first time the problem of a twin cylinder WEC allowed to move in three degrees of freedom. The inclusion of a floating, submerged cylinder as a mechanical reference for power extraction makes this design viable in deep water. With further development of the wave energy industry, it may be anticipated that WECs will follow wind turbines in moving further offshore, making such self-reacting devices more and more relevant \cite{Stiassnie2015}.

Our design procedure initially focused on optimizing device behavior for a damper of constant characteristic in monochromatic waves. At the outset, the heave-only case was considered, presenting a simple situation where a single device (characterized by a size parameter $q$ and a damping parameter $C$), coinciding with the resonant maximum of a freely floating body, outperformed all others. Allowing the device also to sway and roll was seen to introduce additional complexity, and a differentiation was observed between devices operating preferentially in roll/sway and those operating preferentially in heave.

Despite the multiplicity of possible designs when the device is allowed to undergo heave, sway, and roll motions, the monochromatic case presents a clear picture from the standpoint of power absorption: the device closest to heave resonance is found to perform best. This conclusion is an artifact of the idealization represented by the monochromatic theory -- a fact established by the subsequent investigation of WEC performance under an irregular sea.

For our design purposes, a Pierson-Moskowitz spectrum, characterized by wind speed, was chosen to evaluate the designs obtained from the monochromatic case. Under this spectrum, the maxima of absorbed power were found to shift markedly with respect to the monochromatic case, reflecting the potential for detuning to increase captured power in WEC design. Larger values of absorbed power under the design spectrum were found for devices slightly larger and slightly smaller than the monochromatic optimum, raising the question of how to determine device sizing in light of other criteria.

To this end, we have devised some example metrics for grading the sizes of our twin-cylinder WEC. We note that wave energy presents particular difficulties in many respects. While a fixed offshore structure may be designed for survival with very high safety factors, this is inappropriate for oscillating body WECs; by their nature, they must undergo the largest possible motions in order to extract energy. At the same time, device loading should be minimized to avoid fatigue and failure. Taking into account the fact that WECs may be expected to be operational for on the order of 25 years (see Starling \cite{EMEC2009}), and it becomes clear that survival is a paramount issue.  We have presented an example approach to quantitatively evaluate the competing aims of survivability and power extraction within the framework of our floating twin-cylinder device.

To a certain extent all renewable energy technologies, WECs more than most, cannot control their operating conditions, but must work within their environment, subject to the resulting fluctuations of the resource. It must be expected that, like wind turbines, oscillating body WECs will be designed with a ``survival mode", when normal operation cease, and the device changes its characteristics in order to avoid extreme loads. (We might note that overtopping WECs or OWCs (see Section \ref{sec:Introduction}), due to a different working principle and resulting size, will likely have a very different survivability analysis than oscillating body designs.) This may mean increasing the damping, altering the water plane area or mass (see Stallard et al \cite{Stallard2009}), or other approaches (see Coe and Neary \cite{Coe2014}). Due to the nascent state of commercial wave-energy technology, it is difficult to offer concrete design recommendations based on the results for floating twin-cylinders. Our discussion does bear out the fact that a slight over-engineering may be preferable, given the large relative contribution of infrequent, high-energy events to the annual energy budget at many sites, and the demands of survival and robustness. We believe these results to be applicable more broadly to oscillating-body converters, constrained in size as they are by the incident wavelength, indicated by the striking similarities in performance between our twin-cylinder configuration and a single bottom-referenced cylinder.

While many topics of critical importance have not been touched upon -- from moorings to electricity transmission to specifics of power take-off -- we hope to have presented a rather comprehensive picture of the design of a floating twin-cylinder WEC in deep water. There is considerable room for future work, e.g taking into account seasonal resource variability, extreme wave statistics, PTO control, and device cost, as well as detailed studies of reliability under regular operation, where it is hoped that some of the gaps left by this work will be filled in.

\section*{Acknowledgements}
This research was supported by the Israel Science Foundation (Grant 464/13).

\end{document}